\documentclass[sigconf,nonacm,natbib=false]{acmart}
\usepackage{multirow}
\usepackage{filecontents}
\usepackage{xcolor,colortbl}
\usepackage{array}
\usepackage{arydshln}
\usepackage[normalem]{ulem}
\usepackage{longtable}

\copyrightyear{2024}
\setcopyright{rightsretained}
\RequirePackage[
  datamodel=acmdatamodel,
  style=acmnumeric,
  ]{biblatex}

\begin{document}
%TC:ignore
\title[XR for All]{XR for All: Understanding Developers' Perspectives on Accessibility Integration in Extended Reality}
%\title[XR for All]{XR for All: What Makes Accessibility Integration Difficult in Extended Reality?}

\settopmatter{authorsperrow=4}
\author{Daniel Killough}
\email{dkillough@wisc.edu}
\orcid{0009-0002-2623-0528}
\affiliation{%
  \institution{University of Wisconsin-Madison}
  \city{Madison}
  \state{Wisconsin}
  \country{USA}
}

\author{Tiger F. Ji}
\email{reatreify@gmail.com}
\affiliation{%
  \institution{University of Wisconsin-Madison}
  \city{Madison}
  \state{Wisconsin}
  \country{USA}
}

\author{Kexin Zhang}
\email{kzhang284@wisc.edu}
\orcid{0009-0009-4078-8780}
\affiliation{%
  \institution{University of Wisconsin-Madison}
  \city{Madison}
  \state{Wisconsin}
  \country{USA}
}

\author{Yaxin Hu}
\email{yaxin.hu@wisc.edu}
\affiliation{%
  \institution{University of Wisconsin-Madison}
  \city{Madison}
  \state{Wisconsin}
  \country{USA}
}

\author{Yu Huang}
\email{yu.huang@vanderbilt.edu}
\affiliation{%
  \institution{Vanderbilt University}
  \city{Nashville}
  \state{Tennessee}
  \country{USA}
}

\author{Ruofei Du}
\email{me@duruofei.com}
\affiliation{%
  \institution{Google Research}
  \city{San Francisco}
  \state{California}
  \country{USA}
}

\author{Yuhang Zhao}
\email{yuhang.zhao@cs.wisc.edu}
\affiliation{%
  \institution{University of Wisconsin-Madison}
  \city{Madison}
  \state{Wisconsin}
  \country{USA}
}

\renewcommand{\shortauthors}{Killough, et al.}

\definecolor{darkcyan}{rgb}{0.07,0.51,0.77}
\definecolor{gold}{RGB}{255,194,0}
\definecolor{darkgreen}{RGB}{0,100,0}
\newcommand{\TODO}[1]{\textcolor{red}{\textbf{[#1]}}}
\newcommand{\tochange}[1]{\textcolor{red}{\bf{#1}}}%\uwave{}

\newcommand{\yuhang}[1]{{\small\textcolor{red}{\bf [YZ: #1]}}}
\newcommand{\dk}[1]{{\small\textcolor{orange}{\bf [DK: #1]}}}
\newcommand{\daniel}[1]{{\small\textcolor{orange}{\bf [DK: #1]}}}
\newcommand{\rd}[1]{{\small\textcolor{orange}{\bf [RD: #1]}}}
\newcommand{\kz}[1]{{\small\textcolor{blue}{\bf [KZ: #1]}}}
\newcommand{\anonymous}[1]{{\small\color{darkgreen}{\bf[#1]}}}
\newcommand{\change}[1]{{\color{blue}{#1}}}

\begin{abstract}
  % \dk{Seeking to additionally include a note that we are also a team of XR developers to provide a unique perspective in that regard}

%Immersive technologies are poised to revolutionize interactions with recent device launches from companies like Apple and Meta. %\anonymous{adjusted terminology}, 
As immersive technologies enable unique, multimodal interaction methods, developers must also use tailored methods to support user accessibility, distinct from traditional software practices. We interviewed 25 industry extended reality (XR) developers, including  freelancers, startups, midsize, and big tech companies about their motivations, techniques, barriers, and attitudes towards incorporating accessibility features in their XR apps. Our study revealed a variety of challenges, including conflicting priorities between application and platform developers regarding accessibility infrastructure; rapid development culture hindering accessible development; and the lack of accessible interaction design considerations at the ideation, design, and early prototyping stages. As a comprehensive set of XR accessibility guidelines has yet to be established, we also compiled and evaluated a set of accessibility guidelines for 3D virtual worlds and addressed their limitations when applied to XR. Finally, we inform the creation of effective support methods for industry developers.

%Immersive technologies are poised to revolutionize interactions with recent device launches from companies like Apple and Meta. 
% As immersive technologies enable unique, egocentric, multimodal interaction methods, developers must also use unique accessibility methods to support users independent from preexisting methods in traditional software development. 
% To provide recommendations for future developer support methods, our research team of XR developers interviewed 25 industry XR developers representing AR, MR, and VR developers at freelance, startups, midsize / XR engines, and big tech companies on their current perceptions, motivations, techniques, attitudes, and barriers towards developing a11y features in their XR apps. 
% Our study revealed a variety of challenges, including conflicting opinions between application developers and platform developers on accessibility infrastructure, startups' rapid development culture that participants believe prohibits accessible development, and a lack of accessible interaction design considerations at the ideation, design, and early prototyping stages. As a comprehensive set of XR accessibility guidelines has yet to be established, we also compiled and evaluated set of accessibility guidelines for 3D virtual worlds and addressed their limitations when applied to XR platforms. Finally, we establish a framework to inform the creation of effective support methods for industry developers. 
\end{abstract}
%TC:endignore
\begin{CCSXML}
<ccs2012>
   <concept>
       <concept_id>10003120.10011738</concept_id>
       <concept_desc>Human-centered computing~Accessibility</concept_desc>
       <concept_significance>500</concept_significance>
       </concept>
   <concept>
       <concept_id>10003120.10003121.10003124.10010392</concept_id>
       <concept_desc>Human-centered computing~Mixed / augmented reality</concept_desc>
       <concept_significance>500</concept_significance>
       </concept>
   <concept>
       <concept_id>10003120.10003121.10003124.10010866</concept_id>
       <concept_desc>Human-centered computing~Virtual reality</concept_desc>
       <concept_significance>500</concept_significance>
       </concept>
 </ccs2012>
\end{CCSXML}

\ccsdesc[500]{Human-centered computing~Accessibility}
\ccsdesc[500]{Human-centered computing~Mixed / augmented reality}
\ccsdesc[500]{Human-centered computing~Virtual reality}

\keywords{Accessibility, A11y, AR, VR, augmented and virtual reality, extended reality, XR, developer interviews}

% \received{December 2024}
% \received[revised]{August 2025}

\maketitle

\section{Introduction}

% -> important, impactful question everyone would agree on
% -> current solutions
% -> gap
% -> how did we propose to / solve this issue
% -> RQ's
% -> general results 

%\sout{Extended Reality (XR) is revolutionizing how people communicate, interact, and collaborate. \rd{citation? I tend to be}} \sout{\rd{more conservative to use "has the potential to revolutionize"}}
As advances in extended reality (XR) blur the boundaries between physical and virtual worlds, these technologies are reshaping human interaction, communication, and collaboration. XR---an umbrella term encompassing augmented reality (AR), virtual reality (VR), and mixed reality (MR)---utilizes egocentric visual, audio, and haptic feedback to create increasingly immersive experiences.
% The XR market has grown significantly in the past few years, reaching \$29.26 billion USD in 2022 (up from \$18.96 billion in 2021), with Alsop predicting an aggregate worldwide market size of over \$100 billion USD by 2026~\cite{statista2024xrChart}.
% Currently applied in practical fields like education~\cite{kavanagh2017systematic,o2016augmented}, healthcare~\cite{emmelkamp2021virtual,viglialoro2021augmented}, and architecture~\cite{ergun2019architectural,noghabaei2020trend}, .
XR technologies are maturing and growing in popularity in recent years, revolutionizing sectors like education~\cite{kavanagh2017systematic,o2016augmented}, healthcare~\cite{emmelkamp2021virtual,viglialoro2021augmented}, and architecture~\cite{ergun2019architectural,noghabaei2020trend}. 
% \tochange{by reading R2's review again, i think we can simply add stats for XR industry (how widely used is XR on market, how many people own XR devices) and some A11y stats in terms of the distribution of diverse disabilities to highlight how many people are marginalized. would be nice to address this in intro}
% \change{The XR industry reached an estimated \$142.39 Billion USD in 2023 and is projected to grow at a compound annual rate of 32.9\% from 2024 to 2030~\cite{VRMarketSize}, %up to 66\% or higher between 2024 and 2026~\cite{mckinsey2022immersive},
% with more conservative estimates still predicting an over \$500B valuation by 2031~\cite{alsop2024spatial,mordor2024xr,VRMarketSize}. 
Moreover, XR availability is no longer heavily constrained by hardware limitations, as high-resolution VR/MR headsets like the Meta Quest 3S become more affordable~\cite{clique2024quest3s,greener2024metaconnect,truly2024quest3s} and AR apps run on smartphones that consumers already own~\cite{arinsider}. %Although VR head-mounted displays (HMDs) still dominate the market, lightweight head-worn AR smartglasses like the XREAL Air 2 Ultra and Brilliant Labs Frame, evidenced by Snapchat Spectacles unveiled as a restricted developer kit. \tochange{DK: <- include this?}

Although XR platforms offer unique, embodied experiences, %XR adoption growth hinges on technology availability and inclusion~\cite{raji2024business}, and % \tochange{DK: they necessarily mean this in a business sense, but can we replace ``availability'' with ``accessibility'' or is that a false cognate?} 
%of compelling content. %Decreasingly limited by High-fidelity VR devices like the Meta Quest 3 and 3S have become more affordable to consumers, and an increase in Augmented Reality capabilities in modern smartphones have enabled prevalent access to AR applications.
%``
current XR applications pose unique accessibility (a11y~\cite{boia_what_2017}) %\anonymous{"a11y" defined on first use} 
barriers for people with disabilities (PWD)~\cite{wild2024extended}. % compared to \tochange{conventional 2D interfaces}. % a large population of more than 1.3 billion people worldwide ~\cite{WHO2023Disability}.}
%the 1.3 billion people worldwide with ``significant'' disability~\cite{WHO2023Disability}--and the over 2 billion with some form of visual impairment alone~\cite{who2019world,fricke2018global,bourne2017magnitude,flaxman2017global}}. 
%* all of them face different types of issues. consequences could be fatal (e.g. seizures) if not addressed. XR usage is more difficult   
%* XR \-\> can push market distribution if we don’t have exact usage stats among PWD   
%* Trending towards Lightweight e.g. Meta Orion, XREAL air, starting to see with Meta Ray Ban%. 
%\change{those who may benefit from them most}~\cite{citation needed}--- people with disabilities. % e.g. travel barriers, but they also bring significant a11y issues
% \yuhang{too much detail:}
%Although many wearable XR devices use multimodal feedback to immerse their users, most applications are inaccessible to people with disabilities (PWD). 
Unlike traditional 2D software platforms, current XR applications pose unique a11y barriers for people with disabilities (PWD). For example, XR devices do not integrate with assistive technologies that PWD rely on daily, such as XR screen readers for people with visual impairments or bespoke input devices for people with motor impairments. As a result, users with visual impairments may miss essential visual information in a 3D scene~\cite{Waters2001bat}, and users with motor impairments may not be able to complete complex kinesthetic interactions that XR devices require~\cite{mott2020motor}. 
To promote XR a11y, researchers have designed software-~\cite{ji2022vrbubble,khowaja2020augmented,wedoff2019virtual,ye2021paval,zhao2016cuesee} and hardware-based~\cite{choi2018claw,siu2020cane,zhao2018canetroller} assistive technologies, such as audio-based navigation techniques for people with visual impairments in AR~\cite{guo2016vizlens,sato2019navcog3} and VR~\cite{ji2022vrbubble,Waters2001bat} environments. %for example, Ji et al. utilized audio feedback to convey the presence and motion of surrounding avatars to users with visual impairments~\cite{ji2022vrbubble}, while Choi et al. created a custom hand controller for users with very limited hand motion~\cite{choi2018claw}. 
\textit{While effective, most of these technologies remain research prototypes and have not yet seen broad adoption in the XR industry.}% XR applications and platforms.  

% Furthermore, although platforms have the \textit{potential} to incorporate multimodal feedback, many applications lack these features or developers do not implement them. Hand tracking on devices like the Meta Quest do not natively use haptic devices, relying on audio and visual feedback to convey information. 
%For blind and low vision users, deaf and hard of hearing users, those without glasses, people in loud environments, or people simply with the volume turned low, these feedback methods are reduced to one or none. 

% Uses of XR for accessibility itself~\cite{,,}

%Although XR applications focus on immersion, developers have not effectively considered the \change{multitude of people with poor experiences resulting from developers' lack of accessibility implementation~\dk{we don't either in this study}}. 
XR practitioners (e.g., designers, developers, managers) play a vital role incorporating a11y to XR applications---specifically, they must be \textit{motivated} and \textit{able} to implement a11y in their projects. Prior research has highlighted the importance of knowledge, guidance, and organizational structure in supporting a11y integration in consumer products for PWD~\cite{bai2017cost,bi2022pract,miranda2022agile}. While traditional software platforms (e.g., web and mobile applications) have established mature accessibility guidelines~\cite{gameguide, w3web} and automatic accessibility assessments~\cite{achecker, pa11y}, \textit{no standardized guidelines or a11y support methods (e.g., 3D screen reader) exist for XR practitioners to easily implement a11y features}~\cite{johannesson2023early}. %XR development is currently fragmented, with no standard device infrastructure for developers to easily implement a11y features. Further, standardized a11y support methods do not yet exist for XR, such as screen reader for 3D scenes. %\tochange{this is not true, there are many a11y tools in XR, also procedure is not a tool, need to adjust}% nor the ability to recognize the XR screen at a pixel level to evaluate its content.  
%\dk{do we need citations for each of these points, e.g. wobbrock2006analyzing for the fast text input methods section as an example of an equivalent feature in 2D interfaces? or just leave it alone?}.
With XR platforms continually improving, we must understand XR practitioners' current practices and perspectives on a11y, motivating and supporting them to create more accessible XR applications. One recent work evaluates \textit{VR} practitioners' general perspectives on a11y across different development stages ~\cite{wang2025understanding}. However, to derive practical and actionable insights into XR a11y development support that would be broadly accepted and adopted in industry, more thorough research is needed to understand %developers' implementation processes, organizational constraints, and support needs when integrating accessibility features. 
the specific a11y design and development challenges in XR, the support methods (e.g., toolkits, guidelines, \textit{post-hoc} proxies) required by developers, and how these requirements change across organizational contexts (from startup to big tech). 

Our research fills this gap by addressing the following questions: 
\begin{itemize}
    \item[\textbf{RQ1.}] What challenges do XR practitioners face when designing and developing XR a11y features?
    \item[\textbf{RQ2.}] What XR development practices do practitioners follow, such as programming habits, toolkit usage, and testing methods; and how would their practice affect XR a11y integration?
    \item[\textbf{RQ3.}] How applicable are state-of-the-art a11y guidelines to XR, and what support methods (e.g., guidelines, toolkits, platform support) would best enable XR practitioners to integrate a11y features in their applications?
\end{itemize}

To answer these questions, %we interview 25 professional XR developers to understand their needs related to integrating accessibility to their applications. 
we conducted semi-structured interviews with 25 professional XR practitioners to observe their development process via example projects, uncover their perceptions of state-of-the-art a11y guidelines in games and 3D virtual worlds~\cite{apx,gameguide,sig2022guidelines,xbox,oculusvr,w3cxr}, and understand how a11y experts could best support them in creating accessible apps. Our participants represented wide coverage in XR industry, ranging from freelancers and startup employees to big tech companies and XR engine platforms, providing a comprehensive view of XR developers' experiences and needs. Our research revealed unique challenges towards incorporating XR a11y % \kz{add examples}
(e.g., balancing immersion with a11y, text legibility issues, making kinesthetic interactions accessible), uncovered workplace attitudes towards a11y that limit practitioners' agency in developing a11y features (e.g., perceiving a11y as a secondary priority, not considering people with disabilities as target audience, prioritizing financial viability over inclusivity), and highlighted the mismatch between existing a11y guidelines and the unique needs in creating XR experiences (e.g., immersion, spatial audio placement challenges, performance constraints of standalone devices, maintaining competitive integrity in multiplayer apps). %Our guideline evaluation found recommendations like colorblindness support to be widely understood and implementable, but others (e.g., ``Include an option to adjust the game speed''~\cite{cog4-gameguide}) largely inhibit immersion in simulation apps and compromise competitive integrity in multiplayer titles. %These findings highlight unique tensions between developer attitudes and needs of the community, and emphasize the importance of educating developers on the social model of disability~\cite{shakespeare2006social,upias_da_fundamental_1975}. 
Finally, we derive end-to-end implications for actionable a11y integration, from a11y consideration in early development cycle, to toolkits that enable a11y feature implementation, automated a11y checking tools, a11y feature maintenance, data exposure for 3rd party a11y plugins, to PWD engagement in app design and evaluation.  %recommendations for effective services and tools, to best support XR developers in making their apps more accessible. %\change{We contribute these findings} %including recommendations for actionable accessibility support methods, introducing needs for OS- and application-level support, 
%to make XR experiences more accessible to users with disabilities. 

\section{Related Work}
%%% highlight gaps at the end of each subsection
Accessibility in traditional software development has been extensively studied~\cite{bi2022pract,bigham2010accessibility,carter2001web,sodergaard2024enhancing}, with XR recently emerging as a new medium that enables immersive experiences. To contextualize our work, we discuss existing a11y techniques for XR, a11y guidelines for platforms within and beyond XR contexts, and efforts in industry a11y integration.

\subsection{Technology to Enhance XR A11y}
% \yuhang{Add sentence on: "PWD face different challenges in XR, including [citation] and [citation]". Then talk about solutions. Cover empirical or observational work to cover PWD experiences in XR. Check new CHI works on understanding VR A11y (chi24)}
% \tochange{YZ: R1 recommended a list of papers to add about XR accessibility techniques and games. It's also mentioned in the meta review.}
PWD face different challenges in XR, including kinesthetic interactions for wheelchair users~\cite{gerling2020virtual}, unpredictability of VR content increasing concerns for people with photosensitivity~\cite{south2024barriers}, and navigating social VR environments~\cite{collins2024ai}. %identifying and evaluating apps for a11y features~\cite{martinez2024playing}, and audio processing when combining real-world and virtual audio sources~\cite{Chang2024SoundShift} \yuhang{need to rewrite...these examples are not all challenges faced by PWD except for the first one; you are writing about the "tech challenges" not the actual challenges from PWD perspective}.
Existing literature on XR a11y has largely focused on people with visual \cite{heuten2006city,picinali2014exploration,zhao2019seeingvr} and motor \cite{mott2020motor, yamagami2021two} impairments.
For people with visual impairments (PVI), prior work has used 3D audio to translate 2D objects~\cite{heuten2006city} and interfaces~\cite{oliveira2015music} to 3D, convey 3D environments~\cite{balasubramanian2023enable,lokki2005nav, nair2021navstick,nair2022uncovering,walker2006navigation} themselves, and support avatar awareness~\cite{ji2022vrbubble}. %, and simulate real-world scenarios~\cite{}. 
For example,  %~\tochange{DK: R2 mentioned replacing "social distances" with "physical distances" here but Tiger wrote "social" and he's the best person to ask? --- no need, we wrote social distance in that paper, but just follow the review, it's minor} % putting it back to "social" out of spite. Can also just remove the word entirely.
Waters and Abulala \cite{Waters2001bat} simulated echolocation to enable PVI to explore VR environments. %Nair et al. \cite{nair2021navstick, nair2022uncovering} presented a VR exploration technique, allowing PVI to tilt a controller's joystick and receive information of objects in that direction through spatial audio descriptions. %They further iterated on this technique, comparing the directional joystick with other directional interaction methods including a smartphone companion app, echolocating from the player, and a menu-based object exploration technique~\cite{nair2022uncovering}. 
Ji et al.~\cite{ji2022vrbubble} designed an audio-based interaction technique that notified users of surrounding avatars based on social distance. Beyond VR accessibility, Herskovitz et al. \cite{herskovitz2020making} further designed techniques to make mobile AR tasks more accessible to PVI with a series of design prototypes evaluated with 10 blind participants. In addition to audio feedback, researchers also utilized haptics to enable PVI to communicate more effectively in VR~\cite{jung2024accessible}, explore and navigate %real~\cite{sato2019navcog3} and 
virtual spaces~\cite{schloerb2010blindaid}, or interact with VR objects~\cite{jansson1999haptic, tzovaras2009interactive, tzovaras2002design, wedoff2019virtual}. %~\tochange{DK: ideally break this up but low priority + already low on word count + reviewers didn't mention it - it's fine to keep it as it is} 
For example, Jung et al.~\cite{jung2024accessible} combined a series of haptic and audio cues to notify PVI of nonverbal social cues like eye contact, head nodding, and head shaking. Zhao et al.~\cite{zhao2018canetroller} created a wearable VR controller based on a brake mechanism to simulate the white cane interaction for PVI in VR. Siu et al.~\cite{siu2020cane} further improved the controller and provided audio feedback to emulate echolocation. % creating one physical manifestation of Waters and Abulala's work~\cite{Waters2001bat}.

For people with motor impairments (PMI), researchers have explored how motor impairments impact a user's interaction within virtual spaces~\cite{gerling2020virtual, guo2014effects, samaraweera2013latency, yamagami2021two}. For example, Mott et al. \cite{mott2020motor} conducted a survey of 16 PMI to understand what barriers they faced in VR and how VR systems might be improved for those users. They identified issues with both VR interactions and the hardware needed for VR. 
Technologies have also been created to support PMI using XR~\cite{choi2018claw, harada2007voicedraw, lee2019torc}. For example, Choi et al.~\cite{choi2018claw} designed a haptic controller that detected input and generated feedback using motion from only one finger. Harada et al.~\cite{harada2007voicedraw} enabled PMI to use voice input to draw hands-free in VR and created insights for mapping voice input to other continuous motions.

Despite a myriad of assistive technologies developed to facilitate XR a11y, most of these works remain prototypes in the research field without integration into mainstream XR applications or platforms. Although some 2D games like Vampire Survivor~\cite{hall2023vampire} and Loop Hero~\cite{familygaming2024loop} have accessibility features, to our knowledge only a small number of \textit{3D} games, much less XR titles, have incorporated a11y features outside of basic text size or colorblind settings (e.g., The Last of Us Part II~\cite{lastofus}, Mortal Kombat 11~\cite{wbgames2024accessibility}, Age of Empires IV~\cite{worldsedge2024accessibility}).
%God of War: Ragnar{\"o}k~\cite{godofwar},%Some commercial titles have accommodated users with disabilities, either as games specifically made for people with disabilities \cite{BlindSwordsman,ShadeofDoom,AHeroCall} or games for the general public with a notable amount of accessibility features. 
%However, XR applications do not commonly incorporate many accessibility features. 
Therefore, one primary goal of our work is to understand if, how, and why XR developers incorporate a11y into their applications, what factors hinder them to do so, and how a11y techniques might best be translated from research into XR development workflows. 

\subsection{Immaturity of XR A11y Guidelines versus 2D and 3D}
A11y guidelines are important resources that direct developers how to integrate a11y features in their workflows. Standard a11y guidelines have been constructed for various conventional software platforms, such as web \cite{brajnik2009valid, kurniawan2005older, mozilla, w3web} and mobile devices \cite{w3mobile}. For example, the Web Content Accessibility Guidelines (WCAG)~\cite{w3web} have been maintained by the World Wide Web Consortium (W3C) to make websites accessible to people with diverse disabilities. Moreover, W3C also created a set of guidelines for mobile a11y \cite{w3mobile}, addressing mobile-specific concerns such as form factor and potential use in different mobile settings, such as public spaces or in bright sunlight. These guidelines have been used in various studies evaluating mobile platforms \cite{ballantyne2018mobile, androidaccess, wei2018heuristic}. Beyond guidelines for conventional platforms, we focus on discussing state-of-the-art a11y guidelines for the emerging 3D virtual world and XR platforms. 

\subsubsection{Guidelines for 3D Virtual Worlds}
Although both displayed on a 2D screen%\kz{need more clear distinguish of 2d web - 2d virtual world - HMD XR to better contextualize readers}
, guidelines for 2D web applications aren't fully applicable to 3D virtual worlds, such as video games and social virtual communities (e.g., Second Life~\cite{SecondLifeOfficial}).
% \textit{These 3D experiences differ in several key ways: (1) 2D web and mobile applications present content on fixed screens with standard input devices (e.g., keyboard and mouse or touchscreen) where users interact with information in a consistent frame. (2) 3D virtual worlds projected onto 2D screens, like 3D games, represent spatial information and complex navigation, requiring different a11y approaches than traditional 2D web or mobile applications. (3) Immersive XR through head-mounted displays or touchscreens introduce unique challenges including physical movement requirements, spatial disorientation, and embodied interactions that neither 2D nor projected 3D guidelines fully address.}
Dedicated a11y guidelines for virtual worlds have been previously constructed \cite{apx, gameguide, xbox}. For example, the Game Accessibility Guidelines (GAG) \cite{gameguide} provided a comprehensive collection of guidelines separated into disability-based categories. Research efforts have also derived guidelines \cite{mason2022including, mueller2014movement, porter2013empirical, westin2018game} for more specific use cases. For example, Mason et al. \cite{mason2022including} studied challenges faced in movement-based games by wheelchair users, presenting interviewees with eight different game concepts and condensing guidelines to avoid physical barriers. 

\subsubsection{XR-specific Guideline Creation}
XR development is a newer and more fragmented field with no widely-agreed standard. Compared to desktop 3D virtual worlds,
%share some similarities with XR applications, primarily as they both---in a computer graphics context---render 3D applications out to a 2D screen (or, for head-mounted displays, two 2D screens). But 
XR apps are commonly presented on stereoscopic head-mounted displays, embody users into 3D environments from a first person perspective, and allow them to interact with virtual elements via kinesthetic interactions~\cite{gerling2020virtual}. % not just control a character.
%\text%it{(For the rest of this paper, we specify `3D apps' as non-XR titles typically played on a 2D desktop or mobile interface, and `XR apps' as extended reality titles played on an AR-supported mobile device (e.g., Pokémon Go~\cite{pokemongo}) or VR head-mounted display).}
Researchers have come up with general guidelines to support immersive XR experiences \cite{catak2020guideline, fracaro2021chem}. For example, Catak et al. \cite{catak2020guideline} analyzed five VR games to create guidelines according to ``human perception, consistent and intuitive interaction, and reliable navigation.'' 
%as core to VR app design and immersion. 
Fracaro et al. \cite{fracaro2021chem} studied how to design effective and engaging training experiences for chemical operators in VR. However, they do not usually consider a11y.

Recently, both researchers \cite{heilemann2021guidelines, berkeleyAccessibilityUniversal, mott2019design, melbourne} and organizations \cite{xraA11y, usabilityheur, magicleap, oculusvr} have started deriving guidelines for accessible XR experiences. For example, Heilemann et al. \cite{heilemann2021guidelines} surveyed existing a11y guidelines for traditional video games, such as GAG, and compiled a set of guidelines that were directly relevant to VR games and interactions. %However, no evaluation has validated the feasibility of these guidelines in VR. 
The University of Melbourne \cite{melbourne} created a page summarizing some disabilities that cause barriers in VR and listed some relevant papers for each disability. XR organizations have also paid more attention to accessibility guidelines. Oculus \cite{oculusvr}  and Magic Leap \cite{magicleap} also made a set of a11y guidelines to advise developers creating accessible apps on their VR or AR devices. %Similarly,  has provided guidelines for its AR device, focusing on potential hearing, subtitle, visual, and mobility barriers for its user-base. 
Additionally, some a11y interest groups have also created their own sets of considerations for accessible XR design \cite{xraA11y}. %or applied usability heuristics to XR \cite{usabilityheur}, 
%but are typically more limited in scope and do not attempt to cover the entirety of an XR experience. %Alternatively, Fracaro et al. \cite{fracaro2021chem} studied how to design effective and engaging training experiences for chemical operators in VR, but despite the goals of creating guidelines to support a general audience, these works did not consider accessibility for users with disabilities.

\textbf{Despite these ongoing efforts, current XR a11y guidelines are in their infancy, without rigorous validation or broad agreement. Moreover, they mostly focus on accommodating an end user and generally lack technical guidance to developers in how to implement these guidelines.}
Therefore, it is important to obtain the perspective of developers in existing XR a11y guidelines to understand their interpretation, implementation barriers, and needs if applying the guidelines in their projects.

\subsection{Development Support for A11y Integration}
% \yuhang{trim.}
\subsubsection{A11y Integration in Traditional Development}
%A11y integration in industry is not easy. 
Researchers have identified challenges of incorporating a11y into traditional development \cite{bai2017cost, bi2021github, bi2022pract, miranda2022agile}. For example, Bai et al. \cite{bai2017cost} found extensive difficulty when adding user testing for PWD into the web development loop. Moreover, Bi et al. \cite{bi2022pract} interviewed 15 accessibility designers and surveyed 365 participants with accessibility-related work experience worldwide, %to understand what elements of a11y posed barriers to developers and their attitudes towards those problems. They 
revealing substantial variance in a11y development and design quality across developers and deriving suggestions for equalizing these gaps.
% the opportunity costs of adding different kinds of accessibility testing, such as disability simulation for developers and utilizing the perspective of real users with disabilities. They found   was very difficult, especially when 

To better incorporate a11y, development tools have been created to assist developers at various phases of traditional software development, such as software plugins during development \cite{apple, uap} or post-hoc a11y support \cite{bigham2006webinsight, bigham2008webanywhere, achecker, pa11y}. 
Post-hoc accessibility proxies like WebAnywhere \cite{bigham2008webanywhere} and Takagi and Asakawa's transcoding proxy \cite{takagi2000transcoding} improve accessibility by adding enhanced functionality through intermediate layers without requiring developers to modify the original content. Similarly, Zhang et al. developed proxies for mobile platforms that modify the input and output of underlying applications to enhance accessibility of mobile applications \cite{zhang2017interaction}. %\yuhang{what about the examples of post-hoc proxies? Refer to SeeingVR RW.} 
Automatic a11y checking tools also exist, such as Accessibility Checker \cite{achecker} and Pa11y \cite{pa11y}, which analyze a website to ensure compliance with web a11y guidelines. % like WCAG and the Americans with Disabilities Act (ADA)~\cite{ADAfactsheet,ADA}. %, informing developers of any potential fixes to the HTML or Javascript. 
However, less research or tools have focused on a11y integration for XR applications. 

\subsubsection{Challenges and Support in XR Development}
While not a11y focused, some research explored the challenges in XR development as a whole \cite{borsting2020towards, krau2021collab, krauss2022elements, murphyhill2014cowboy}. For example, Ashtari et al. \cite{ashtari2020xr} examined the barriers to becoming an XR developer, revealing the significant challenges in locating learning resources and keeping up with evolving platforms and tools. Krau{\ss} et al. \cite{krau2021collab} approached the collaborative aspect of XR development %byinterviewing developers about their roles and engagement with collaborators 
and found that developers tended to take on multiple roles and face many challenges in collaboration, particularly when communicating XR prototypes. To support XR development, a variety of tools have been created, %usually aimed at simplifying or automating common functionalities \cite{normcore, xritool}
such as the Normcore toolkit that supports automatic multiplayer functionality in VR \cite{normcore} and Unity's XR Interaction Toolkit that enables XR controller and camera setup %cross platforms %compatibility with various devices, 
and basic VR interactions \cite{xritool}. %For AR, Unity's UI Accessibility Plugin \cite{uap} also allows developers to attach a component to 2D UI elements to enable screen reader support on smartphones and computers.
Nebeling \cite{nebeling2022tools} also reviewed recent XR development tools, highlighting their increasing approachability for designers and developers familiar with non-XR tools.

Despite extensive research on accessibility techniques for XR, there has been limited investigation into whether and how XR developers may integrate a11y techniques in practice. 
SeeingVR~\cite{zhao2019seeingvr} provided a VR developer toolkit comprising 14 low vision tools to help developers easily incorporate these a11y features in their applications. However, this work mainly focused on the creation and evaluation of the toolkit itself without deeply understanding developers' experiences and needs. % More in-depth investigations are needed to explore XR developers' experiences, barriers, and needs in a11y integration, thus informing the design of most suitable support to motivate and facilitate this process.
The only exception is Wang et al.'s work~\cite{wang2025understanding}, who conducted interviews and surveys with VR practitioners. They found that VR practitioners with experience in traditional software development had a better understanding of a11y needs and identified challenges including hardware limitations, insufficient professional knowledge, and competing development priorities. Compare to this work, we dive deeper into the impact of unique XR characteristics and more thoroughly consider practitioners' design and development needs from different practical perspectives (e.g., toolkits, guidelines) to inspire future XR a11y support methods. Specifically, we consider the broader XR ecosystem including VR, AR and MR development, reveal a11y feature design challenges and tension due to the unique XR nature (e.g., immersion, kinesthetic interactions), examine the impact of different organizational contexts (from freelancers to big tech) on practitioner's a11y decision and resources, and discuss their preferences and concerns in various support methods that could bridge the gap in a11y integration. Moreover, beyond interviews, we also asked XR practitioners to examine existing XR-relevant a11y guidelines and discuss their interpretation, potential use, and concerns with the guidelines for future improvement. %Furthermore, very limited research and support tools have focused on a11y integration into XR development. %the challenges, practices, and tool usage of XR developers in accessibility integration. 

\section{Method}
% \dk{overview of the study (or before)}
% \yuhang{The goal of the study \& research questions -- be specific. clarify 3 aspects you want to cover}
% \yuhang{add a couple of sentences to summarize the goal of our study. "The goal of our study is to xxx"; highlight the three RQs.}
The goal of our study is to deeply investigate XR a11y integration barriers and needs from the practitioners' perspective, understanding their challenges in a11y feature design and development (RQ1), the impact of their current development practices on a11y implementation (RQ2), and their perception, potential use, and concerns with applying relevant a11y guidelines to XR (RQ3).
To achieve this goal, we conducted semi-structured interviews with 25 XR practitioners across diverse organizational contexts, discussed their experiences, observed their development workflows, and evaluated their interpretations of existing XR-relevant a11y guidelines. Our research aimed to uncover actionable insights to bridge the gap between a11y research and industry practice in XR, inspiring effective support methods for universal XR accessibility. 

\subsection{Participants}
We recruited 25 XR developers/designers (23 male and 2 female) with diverse XR development experiences. Twenty-four participants had rich XR development experience ranging from two to nine years ($mean=4.92$, $SD=1.94$), while the last (P17) had less than one year experience but worked on an a11y team at a major XR game engine. Our participants also had experiences in a wide range of XR industry environments, from freelancers working on individual personal projects to large companies working on many projects with diverse teams. 
Specifically, we recruited participants from four levels of employment environments, which can significantly affect their a11y implementation decision: (1) \textit{Freelance}: individuals working for a particular client until the scope of that project is fulfilled; %or those self-describing as a ``freelancer'' or ``contractor''. 
(2) \textit{Startup}: a newly created company that is in early stages of development but aims to grow rapidly and scale its business model, e.g., \textit{VRChat Inc.}; 
%independent companies of multiple employees with specified roles, but one developer often shares multiple responsibilities. Startups produce a specific product (or line of products) but lack established a11y or legal teams.
(3) \textit{Midsize}: a company that falls between a startup and a large corporation in terms of size and revenue, which has more established system than startup but offers growth opportunity with slower pace, e.g., \textit{Unity Software Inc.}; and (4) \textit{Big tech}: established, large technology companies with mature, dedicated a11y and legal teams, e.g., \textit{Microsoft}.

% ``startup'', ``midsize'', and ``big tech.''
% \begin{itemize}
%     \item We consider ``freelance'' to be 
%     \item We consider ``startups'' to be %~\tochange{DK: Do we need a more formal definition than this or good enough? - not worry about it now since it's not mentioned by reviewers; but can revisit in camera ready if accepted.}
%     \item We consider ``midsize'' to be a catch-all term between mature companies, game engines (e.g. Unity, Unreal Engine), universities, and medical facilities. Midsize companies shared some characteristics between startups and big tech but had their own restrictions, like a budding a11y team, as outlined further in Section~\ref{findings}.
%     \item We consider ``big tech`` to be established, large technology companies, e.g. ``MAMAA'' (Meta, Apple, Microsoft, Amazon, Alphabet), or similarly sized companies. Big tech companies often have mature, dedicated a11y and legal teams. 
% \end{itemize}

We first recruited participants using a convenience sampling method through general recruitment on AR and VR forums (e.g., Reddit, Unity). Based on the distribution of our participants, we then conducted expert purposive sampling~\cite{tongco2007purposive,tremblay1957key} with a more targeted recruitment message to developers on LinkedIn, contributors to AR or VR-related projects on GitHub with publicly-available email addresses, and personal connections. %(e.g., more participants from larger companies and to round out our guideline evaluation numbers). 
% We start with general recruitment on forums. Then based on distribution of participant background, we did more targeted recruitment on LinkedIn based on participant background with a more targeted recruitment message. 
%~\tochange{X participants},
% , LinkedIn%~\tochange{X participants}
% , and personal connections %~\tochange{X participants}
% We also contacted contributors to AR or VR related GitHub projects with publicly available emails%~\tochange{X participants}
Participants were eligible if they were at least 18 years old and had at least 1 year of industry experience in XR development.
As a result, 17 participants were recruited through online methods, and eight were from personal connections.  %, \change{given that they may be more likely to screenshare direct} \change{code examples with us}. 
We detail participants' demographic information, including individuals' years of experience and preferred platform(s), in Table~\ref{table:participants}. %Though all participants gave valuable insight, we reached saturation for freelance and startup perspectives after approx. 20 participants. We recruited a few additional participants to round out the data for big tech and midsize companies.

\begin{table*}[ht]
\centering
\small
\caption{Participant Demographic Information, including gender, size of employment environments, years of XR development experience and classified roles, primary development platforms, and prior a11y development experiences.  %Participants with multiple ``Company Size'' listings shared insight from experiences as employees at multiple company sizes, for example as a former employee at a midsize game engine who now is a freelancer. Company sizes are ordered with largest company size first getting smaller, i.e. big tech is greater than midsize > startup > freelance. The ``XR Dev Exp \& Primary Roles'' column notes the duration in years XR developers have experienced and what their primary roles or responsibilities are. Roles are listed with any management positions first, then any design roles, research roles, then developer, if applicable. Primary Dev Platform(s) notes the top development platforms or integrated development environment (IDE) used. If a developer mentioned two platforms as tied, they are both listed. P17's platform was redacted for privacy as it lists their direct place of employment and they work on a small team. The final column outlines developers' prior accessibility experience, categorized into the type of accessibility impairment their features or solutions addressed (visual, motor, cognitive, or speech \& hearing). We also indicate whether participants have experience creating XR a11y tools, e.g., P3's mobile AR screenreader.} %\tochange{YZ: adjust table width; avoid going beyond the margin}
}
\Description{A table showing abridged demographic information of the participants. From left to right, column titles are PID, Gender, Company Size, XR Development Experience \& Primary Roles, Primary Dev Platforms, and Prior A11y Experience. In the first and second columns, participants are numbered 1 through 25, with P2 and P10 listed as Female and the rest Male. For company size, participants have overlapping experience, but the following participants are listed as big tech: P3, P5, P15, P19, P21, P23, P24, and P25; the following participants are listed as midsize: P7, P9, P10, P14, P16, P17, and P24. The following participants are listed as startup: P1, P2, P4, P6, P8, P11, P12, P13, P16, P18, P20, P21, P22, P23. The following participants are listed as freelance: P1, P2, P4, P5, P6, P8, P11, P14, P22, P23. In the fourth column, XR Development Experience \& Primary Roles, participants are listed from less than 1 year for P17, on an a11y team at engine, then 2 years for P9, and up to 9 years for P8. The most participants have 4 to 6 years of experience. All participants except P12 and P18 have developer listed, who have designer listed instead. These participants only have developer listed: P2, P5, P9, P10, P11, P13, P15, P16, P17, P19, P21, and P24. Other participants have designer, researcher, or some sort of management role listed in addition to developer. In the fifth column, Primary Dev Platform(s), Unity is overwhelmingly listed, with few other options, being: BabylonJS Playground for P3; Glitch for P7 and P9; Custom Engine AND Unity for P8; Unreal Engine for P11; PlayCanvas for P14; P17's place of employment (redacted); and both Custom Engine AND Unreal for P25. All other rows just list Unity. The final column, Prior A11y Exp, lists either ``no” or “yes''. If yes, the type of a11y impairment is listed or simply ``XR tools''.  With some overlapping experience, 20 participants are listed as ‘yes’, with 10 participants listed with visual experience: P2, P3, P4, P7, P13, P15, P19, P20, P23, and P24. 7 participants listed with motor experience: P1, P2, P4, P13, P19, P21, and P23. 5 participants listed with cognitive experience: P1, P7, P16, P22, and P25. 4 participants listed with speech and hearing experience: P5, P12, P23, and P25. 4 participants listed with XR tools experience: P3, P6, P15, and P16. 5 participants are listed as no a11y development experience: P8, P9, P10, P11, and P18.}
\resizebox{\textwidth}{!}{\begin{tabular}{@{}lcllll@{}}
\toprule
\textbf{PID} & \textbf{Gender} & \textbf{Company Size Exp} & \textbf{XR Dev Exp \& Primary Role(s)} & \textbf{Primary Dev Platform(s)} & \textbf{Prior A11y Exp} \\
\midrule
P1 & Male & startup, freelance & 5y: designer, writer, developer & Unity & XR motor, cognitive \\
P2 & Female & startup, freelance & 6.5y: developer & Unity & XR visual, motor \\
P3 & Male & big tech & 4y: manager, researcher, developer & BabylonJS Playground & XR visual; XR tools \\
P4 & Male & startup, freelance & 4y: manager, designer, developer & Unity & XR visual, motor \\
P5 & Male & big tech, freelance & 3y: designer, animator, developer & Unity & yes; XR speech, hearing (s\&h) \\
P6 & Male & startup, freelance & 2.5y: manager, developer & Glitch & webXR tools \\
P7 & Male & midsize & 5y: researcher, developer & Unity & XR visual, cognitive \\
P8 & Male & startup, freelance & 9y: designer, developer & Custom Engine; Unity & None \\
P9 & Male & midsize & 2y: developer & Glitch & None \\
P10 & Female & midsize & 3y: developer & Unity & None \\
P11 & Male & startup, freelance & 5y: developer & Unreal & None \\
P12 & Male & startup & 3.5y: designer & Unity & XR hearing \\
P13 & Male & startup & 2y: developer & Unity & XR motor \\
P14 & Male & midsize, freelance & 3y: researcher, developer & PlayCanvas & XR visual \\
P15 & Male & big tech & 5.5y: designer, developer & Unity & XR visual; XR tools \\
P16 & Male & midsize, startup & 4y: designer, researcher, developer & Unity & XR cognitive; XR tools \\
P17 & Male & midsize & <1y: developer & \textit{Redacted XR Engine} & \textit{On a11y team at XR engine} \\
P18 & Male & startup & 6y: executive, designer, researcher & Unity & None \\
P19 & Male & big tech & 8y: developer & Unity & XR visual, motor \\
P20 & Male & startup & 7y: team lead, developer & Unity & XR visual \\
P21 & Male & big tech, startup & 5y: developer & Unity & XR motor \\
P22 & Male & startup, freelance & 5y: team lead, developer & Unity & XR cognitive \\
P23 & Male & big tech, startup, freelance & 8y: designer, developer & Unity & XR visual, motor, speech \\
P24 & Male & big tech, midsize & 7y: developer & Unity & XR visual \\
P25 & Male & big tech & 5y: team lead, developer & Custom Engine; Unreal & XR cognitive, s\&h \\
\bottomrule
\end{tabular}}
\label{table:participants}
\end{table*}
%TC:endignore

\subsection{Procedure}
We conducted a semi-structured interview via Zoom with each participant. %Studies were conducted in English between October 2022 and July 2023. 
Each study lasted approximately two hours, %split over %($min=1h40m, max=2h54m, median=2h2m$) 
%one or two sessions, 
and we compensated participants at \$50/hr. 
% Studies were conducted October 2022 to July 2023. 
Each interview included four phases: (1) Demographic and past development experience in XR and A11y; (2) Specific examples of past XR and/or a11y projects; (3) Evaluating existing a11y guidelines that can be applied to XR; and (4) Perspectives and challenges of XR a11y implementation support, such as 3rd party developer toolkits and \textit{post-hoc} a11y proxies. We detail each section as follows:

\textit{Demographic \& Past XR/A11y Development Experience.}
We first asked participants' demographic and background information, including gender identification, education formally (e.g., university) and informally (e.g., self-taught), work organization, job position, and how long they have worked as XR developers. We then asked about their XR development experience%, such as the XR devices they developed for; the development platforms, languages, and/or frameworks they used; and a brief description of the types of XR applications they have developed (e.g., XR games, tools, simulators, social applications). 
and to describe the typical design and development cycle of an XR project in their organization. %including their design, collaboration, and testing practices. %such as how many people were involved in an XR project and what their roles and responsibilities were. We also asked how participants designed XR features, collaborated with one another, and tested XR features.

\textit{Specific XR/A11y Projects.}
To understand XR developers' a11y experience and attitudes toward a11y integration in XR projects, we asked participants about their general a11y knowledge and experiences across different platforms (e.g. web, mobile, desktop). We then asked whether they had XR-specific a11y experience; and if so, to detail the features developed and the development cycle, from motivations, to design, tools used, and how they tested those features. %; their motivations to develop such features; the target audiences of the application; and their design, development, and testing processes. We also asked what resources they used to develop these features (e.g., toolkits, packages, or guidelines), how they found those resources, and how useful those resources were. 
If not, we asked what prevented them from considering or incorporating a11y features in their XR apps. %have any experience developing accessibility features for XR applications, we asked whether they had considered people with disabilities as part of their target audience and what prevented them from considering or incorporating any accessibility features. 
If a participant had experience developing both XR and non-XR a11y features, we asked them to discuss the differences between implementing a11y features for XR versus non-XR applications. 
% We then asked all developers about their attitudes towards accessibility integration in XR projects and the barriers they faced. 

We then asked participants to screen-share one XR project, understanding their XR development practices and garnering insight towards potential a11y support methods that best fit their existing practices. During this phase, we observed participants' XR development process, such as how they set up their development environment, how they organize their scene hierarchy, and what naming practices they follow when creating virtual elements. We also asked about participants' use of 3rd party packages and their affinity towards them. %3rd party packages, including what packages they had installed (if any).

\textit{XR A11y Guideline Evaluation.}
We then presented a selection of existing XR a11y guidelines and asked participants to discuss their interpretations, potential implementation plans, and concerns. To enable this phase of the study, we collected a set of 20 guidelines that support various disabilities (i.e., visual, motor, cognitive, speech, and hearing impairments) by reviewing 12 guideline resources. Section \ref{guidelines} elaborates our guideline selection details.  %talked about their interpretation of these guidelines, and discussed how they would implement the guidelines in their projects and the challenges they anticipated. 
% We collected 20 guidelines based on state-of-the-art research and industry practices, and we organized them based on four types of disabilities: visual, motor, cognitive, and speech \& hearing. We further detail the guideline selection process in \S~\ref{guidelines} and present a summary of the guidelines in Table~\ref{table:guidelines}.
%Due to time constraints,

Due to time constraints of the study, we asked each participant to select and discuss only one group of guidelines that focused on one specific type of disability. To enable participants to discuss the guidelines that they prefer to and also ensure all groups of guidelines are equally discussed across participants, we adopted the following guideline assignment strategy: We first asked each participant to choose two disability types that they were interested in; we then selected one disability from these two options for the participant as their final group for discussion.
Participants largely chose particular disability groups due to past experience developing solutions for that group (e.g., aligning with P7's target audience); they had a friend or personal identity with that group (e.g., P6 having a colorblind friend%, P9 experiencing ADHD~\tochange{DK: is this appropriate language?-fixed}
); or they were interested in building a11y solutions for that disability group (e.g., P2 was interested in audio-only XR experiences).

\textit{Perspectives and Challenges on XR a11y integration.} We closed the interview by asking participants' overall perspectives on integrating a11y features into their projects. 
Specifically, we asked about the time and effort they felt viable to dedicate to a11y in their XR projects, the greatest barriers to a11y implementation, and their preferred assistance method (e.g., guidelines, developer toolkits, \textit{post-hoc} plugins or proxies like an XR screen reader) to best support a11y integration at different development phases. % including at what development phase such assistance could be best applied. %Finally, as assistance could be provided either during development or through post-hoc plugins from third parties \cite{zhao2019seeingvr}, we asked participants which method they would prefer and why. 
%We also asked developers' thoughts on supporting 3rd party modification of their XR applications with accessibility tools. 
%We closed the interview asking how much effort developers would want to put towards supporting a11y features, and how much they expect 3rd party support methods to contribute. %how much effort they would want to put towards creating metadata for post-hoc accessibility support, such as screen readers; how much effort should be provided by third-parties towards supporting accessibility; and whether and why they would be willing to adopt new development practices to integrate accessibility.

\subsection{Apparatus: Guideline Selection}
\label{guidelines}
We collected existing a11y guidelines that can be applied to XR for developers to review, thus understanding their interpretations, challenges, and needs for applying the guidelines. While there is still no commonly-agreed, mature XR a11y standards, we collected existing preliminary guidelines distilled by researchers and accessibility organizations. %to understand developers' interpretations and needs for guidelines that can best support accessibility integration. 
To achieve a comprehensive collection of a11y guidelines, we conducted a systematic online search using combinations of the keywords ``Augmented Reality (or AR),'' ``Virtual Reality (or VR),'' ``Mixed Reality (or MR),'' ``Extended Reality (or XR),'' with ``accessibility guidelines,'' retrieving relevant results from the first five pages of Google and Google Scholar. We also decided to include ``Game'' in our search, as: (1) dedicated guidelines for XR are limited; and (2) there is significant overlap in design paradigms between XR applications and 3D games and the tools to make them (e.g., Unity)~\cite{gunkel2021immersive}---
particularly as VR applications and VR games introduce unique a11y challenges beyond traditional 3D games~\cite{heilemann2021guidelines}.
% , we also included ``Game'' in our search. %We need VR-specific guidelines because VR introduces unique accessibility challenges (e.g., motion sickness, spatial navigation) not covered by traditional guidelines, and existing game guidelines lack actionable VR recommendations. This aligns with Heilemann et al.'s call towards tailored VR accessibility standards to address unmet industry needs ~\cite{heilemann2021guidelines}. 
We further filtered the search results by removing resources that were clearly not relevant to XR or 3D games. 

% *(Note: Removed citations/references for conciseness, but they can be re-added if needed.)* 
% \yuhang{refer to heilemann2021guidelines to explain why focusing on VR relevant guidelines.}
% \tochange{explain the relationship between game and XR, and why game guidelines are highly relevant to XR -- 2AC}. Given 

% At the time of the study (October 2022), 
Our search yielded 12 guideline resources \cite{apx, androidaccess, gameguide, uap, heilemann2021guidelines, sig2022guidelines, magicleap, xbox, oculusvr, melbourne, w3mobile, w3cxr}. To narrow the scope and identify a small set of guideline examples for participants to examine, we focused on relatively mature, comprehensive guideline resources. %only focused on a narrow range of disabilities (e.g., \cite{uap, magicleap}), %\tochange{R1 is wondering why this is the case}
%\tochange{if this sentence is potentially a point of contention and tiger just didn't see them, then should we just take them out? they didn't come up in our initial search, so don't mention it and see if the reviewer wants to add it?}
We excluded guidelines that primarily provided conceptual information without specific a11y issues and implementation recommendations (e.g., \cite{melbourne}), %could not be easily applied to 3D games or XR content %\change{(to prompt developers to focus on XR titles' 3D context in their responses)} 
%(e.g., \cite{androidaccess}), 
or guidelines that were a subset of more comprehensive resources already included in our analysis (e.g., \cite{heilemann2021guidelines, w3mobile}). 
As a result, we narrowed down to six resources, including the Game Accessibility Guidelines (GAG) \cite{gameguide}, the W3C XR Accessibility User Requirements (W3CXR) \cite{w3cxr}, Designing Accessible VR instructions from Meta Quest (Quest) \cite{oculusvr}, the Accessible Player Experiences guidelines (APX) \cite{apx}, the Xbox Accessibility Guidelines (Xbox) \cite{xbox}, and the IGDA GASIG Top Ten guidelines (IGDA) \cite{sig2022guidelines}. We then synthesized a common subset across most resources. Specifically, we used GAG as an anchor, matched its guidelines with other resources, added new guidelines from other resources, and labeled duplicate guidelines. %and finally selected most commonly mentioned five guidelines for each type of disability. 
We finally selected and organized guidelines based on disability type---motor, visual, cognitive, and speech \& hearing disabilities---and selected the most commonly mentioned five guidelines for each group. We merged the guidelines for speech and hearing disabilities as both emerged as highly relevant to communication, and they had smaller number of guidelines than other disability types. (Some resources already merged these guidelines; for example, the GAG noted
text chat as benefiting d/Deaf and non-verbal users together~\cite{sh2-gameguide}.)
%We narrowed down our results to the five most reoccurring guidelines per group across all resources. 
%We acknowledge that the guidelines selected still do not fully translate to all XR applications due to the immaturity of XR platforms and how these a11y guidelines were written, but we tried to narrow down to the most relevant guidelines and use them as probes to inspire future XR a11y guidelines. \tochange{any further justification needed? or too much?}
% Our search resulted in four disability groups with each group consisting of five most reoccurring guidelines across all resources. 
We include a full list of these guidelines with their resources in Table~\ref{table:guidelines}. %  have cited the original wording for each guideline. %\tochange{DK: Should we try to go deeper here if they're specifically looking for us to "translate these into validated XR-specific guidelines. The findings reveal that developers found these guidelines lacking in XR context - as noted on page 16: `guidelines need more XR-specific and a11y-specific context`' and some guidelines like timing controls seemed "possible to implement but not applicable to many XR apps."? --- not necessary; and there's no way for us to go deeper as this is the method section and we did not reveal any findings yet. Could bring this up in Discussion (e.g., limitation) when you see fit.}
%\tochange{add that we acknowledge that the guidelines still do not fully translate to XR due to the immaturity of XR a11y guidelines,  and they may also vary for different XR apps, we tried to narrow down to the most relevant ones and use them as probes to inspire future XR a11y guidelines. --- re R2}}

\newcolumntype{P}[1]{>{\arraybackslash}m{#1}}
%\newcolumntype{L}[1]{>{\arraybackslash}m{#1}}

\newcommand{\tabitem}{~~\llap{\textbullet}~~}
%TC:ignore
\begin{table*}[h]
  \caption{Selected XR a11y Guidelines, grouped by disability type (Visual, Motor, Cognitive, and Speech \& Hearing). %Disability types are listed in (and guidelines are numbered in) the same order as presented to participants. 
  Guidelines are abbreviated as Visual---Vis, Motor---Mot, Cognitive---Cog, and Speech \& Hearing---SH. Original wording for each guideline from different sources are linked via the references next to each guideline. 
  %As used as an anchor, references to the GAG---Game Accessibility Guidelines~\cite{gameguide} are listed first, then all subsequent resources are listed in alphabetical order: APX---the Accessible Player Experiences guidelines~\cite{apx}; IGDA---the IGDA GASIG Top Ten guidelines~\cite{sig2022guidelines}; Quest---Meta Quest's Designing Accessible VR~\cite{oculusvr}; W3CXR---the W3C XR Accessibility User Requirements~\cite{w3cxr}; Xbox---the Xbox Accessibility Guidelines~\cite{xbox}. \change{Some guidelines cite the same source in cases where the source either lists a large number of guidelines and does not have section headers (e.g., \cite{sig2022guidelines,top10-sig2022guidelines}) or includes multiple related guidelines (e.g., \cite{mot2-apx,mot1-xbox}). These guidelines are still divided by their GAG~\cite{gameguide} anchor and citations are listed multiple times below.} \tochange{too much information or good as-is?}
  }
  \small
  \Description{A table presenting selected XR Accessibility Guidelines, categorized by disability type. The table has two columns: "Disabilities" and "Guidelines". The Disabilities column lists four categories: Visual Impairments, Motor Impairments, Cognitive Impairments, and Speech & Hearing Impairments. The Guidelines column provides five guidelines for each disability category, numbered and abbreviated (e.g., Vis-1, Mot-2, Cog-3, SH-4). Each guideline is followed by source references in brackets. Visual Impairment guidelines include:
1. Ensuring essential information isn't conveyed by color alone; 2. Allowing resizable interfaces; 3. Using easily readable default fonts; 4. Providing audio description tracks; and 5. Offering pre-recorded voiceovers for all text. Motor Impairment guidelines include: 1. Allowing remappable/reconfigurable controls; 2. Supporting multiple input devices; 3. Avoiding reliance on motion tracking of specific body types; 4. Ensuring simple controls or providing simpler alternatives; and 5. Avoiding essential precise timing in gameplay. Cognitive Impairment guidelines include: 1. Avoiding flickering images and repetitive patterns; 2. Reinforcing text information with visuals and/or speech; 3. Using symbol-based chat; 4. Including options to adjust game speed; and 5. Providing options to hide non-interactive elements. Speech & Hearing Impairment guidelines include: 1. Providing captions or visuals for significant background sounds; 2. Supporting text chat alongside voice for online multiplayer; 3. Ensuring no essential information is conveyed by sounds alone; 4. Providing separate volume controls for effects, speech, and background/music; and 5. Allowing customizable subtitle/caption presentation.}
  \begin{tabular}{P{2.4cm}P{12cm}}
    \toprule
    \textbf{Disabilities} & \textbf{Guidelines}\\
    \toprule
    \multirow{5}{5em}{\textbf{Visual \\ Impairments}} 
      & \tabitem \textit{Vis-1.} Ensure no essential information is conveyed by a color alone. (\href{https://gameaccessibilityguidelines.com/ensure-no-essential-information-is-conveyed-by-a-colour-alone/}{GAG}~\cite{vis1-gameguide}, \href{https://accessible.games/accessible-player-experiences/access-patterns/distinguish-this-from-that/}{APX}~\cite{vis1-apx}, \href{https://igda-gasig.org/get-involved/sig-initiatives/resources-for-game-developers/sig-guidelines/on-visual-disabilities/}{IGDA}~\cite{visual-sig2022guidelines}, \href{https://developer.oculus.com/resources/design-accessible-vr-display/\%23color}{Quest}~\cite{vis1-meta}, \href{https://learn.microsoft.com/en-us/gaming/accessibility/Xbox-accessibility-guidelines/103}{Xbox}~\cite{vis1-xbox})\\
      & \tabitem \textit{Vis-2.} Allow interfaces to be resized. (\href{http://gameaccessibilityguidelines.com/allow-interfaces-to-be-resized}{GAG}~\cite{vis2-gameguide}, \href{https://accessible.games/accessible-player-experiences/access-patterns/personal-interface/}{APX}~\cite{vis2-apx}, \href{https://igda-gasig.org/get-involved/sig-initiatives/resources-for-game-developers/sig-guidelines/on-visual-disabilities/}{IGDA}~\cite{visual-sig2022guidelines}, \href{https://www.w3.org/TR/xaur/\%23interaction-and-target-customization}{W3CXR}~\cite{vis2-w3cxr}, \href{https://learn.microsoft.com/en-us/gaming/accessibility/Xbox-accessibility-guidelines/112}{Xbox}~\cite{vis2-xbox})\\
      & \tabitem \textit{Vis-3.} Use an easily readable default font. (\href{http://gameaccessibilityguidelines.com/use-an-easily-readable-default-font-size}{GAG}~\cite{vis3-gameguide}, \href{https://accessible.games/accessible-player-experiences/access-patterns/clear-text/}{APX}~\cite{vis3-apx}, \href{https://developer.oculus.com/resources/design-accessible-vr-captions/}{Quest}~\cite{vis3-meta}, \href{https://learn.microsoft.com/en-us/gaming/accessibility/Xbox-accessibility-guidelines/101}{Xbox}~\cite{vis3-xbox})\\
      & \tabitem \textit{Vis-4.} Provide an audio description track. (\href{http://gameaccessibilityguidelines.com/provide-an-audio-description-track/}{GAG}~\cite{vis4-gameguide}, \href{https://accessible.games/accessible-player-experiences/access-patterns/second_channel/}{APX}~\cite{vis4-apx}, \href{https://www.w3.org/TR/xaur/\%23gestural-interfaces-and-interactions}{W3CXR}~\cite{vis4-w3cxr}, \href{https://learn.microsoft.com/en-us/gaming/accessibility/Xbox-accessibility-guidelines/111}{Xbox}~\cite{vis4-xbox})\\
      & \tabitem \textit{Vis-5.} Provide pre-recorded voiceovers for all text, including menus and installers. (\href{http://gameaccessibilityguidelines.com/provide-full-internal-self-voicing-for-all-text-including-menus-and-installers}{GAG}~\cite{vis5-gameguide}, \href{https://igda-gasig.org/get-involved/sig-initiatives/resources-for-game-developers/sig-guidelines/on-visual-disabilities/}{IGDA}~\cite{visual-sig2022guidelines}, \href{https://developer.oculus.com/resources/design-accessible-vr-design/}{Quest}~\cite{vis5-meta}, \href{https://www.w3.org/TR/xaur/\%23immersive-semantics-and-customization}{W3CXR}~\cite{vis5-w3cxr}, \href{https://learn.microsoft.com/en-us/gaming/accessibility/Xbox-accessibility-guidelines/106}{Xbox}~\cite{vis5-xbox})\\
    \midrule
    \multirow{5}{5em}{\textbf{Motor \\ Impairments}} 
      & \tabitem \textit{Mot-1.} Allow controls to be remapped/reconfigured. (\href{http://gameaccessibilityguidelines.com/allow-controls-to-be-remapped-reconfigured}{GAG}~\cite{mot1-gameguide}, \href{https://accessible.games/accessible-player-experiences/access-patterns/same-controls-but-different/}{APX}~\cite{mot1-apx}, \href{http://igda-gasig.org/get-involved/sig-initiatives/resources-for-game-developers/sig-guidelines/}{IGDA}~\cite{top10-sig2022guidelines}, \href{https://developer.oculus.com/resources/design-accessible-vr-controls/}{Quest}~\cite{mot1-meta},  \href{https://www.w3.org/TR/xaur/\%23xr-controller-challenges}{W3CXR}~\cite{mot1-w3cxr}, \href{https://learn.microsoft.com/en-us/gaming/accessibility/Xbox-accessibility-guidelines/107}{Xbox}~\cite{mot1-xbox})\\
      & \tabitem \textit{Mot-2.} Support more than one input device. (\href{http://gameaccessibilityguidelines.com/support-more-than-one-input-device}{GAG}~\cite{mot2-gameguide}, \href{https://accessible.games/accessible-player-experiences/access-patterns/flexible-controllers/}{APX}~\cite{mot2-apx}, \href{http://igda-gasig.org/get-involved/sig-initiatives/resources-for-game-developers/sig-guidelines/}{IGDA}~\cite{top10-sig2022guidelines}, \href{https://www.w3.org/TR/xaur/\%23various-input-modalities}{W3CXR}~\cite{mot1-w3cxr}, \href{https://learn.microsoft.com/en-us/gaming/accessibility/Xbox-accessibility-guidelines/107}{Xbox}~\cite{mot1-xbox})\\
      & \tabitem \textit{Mot-3.} Do not rely on motion tracking of specific body types. (\href{https://gameaccessibilityguidelines.com/do-not-rely-on-motion-tracking-of-specific-body-types/}{GAG}~\cite{mot3-gameguide}, \href{https://accessible.games/accessible-player-experiences/access-patterns/flexible-controllers/}{APX}~\cite{mot2-apx}, \href{https://developer.oculus.com/resources/design-accessible-vr-controls/}{Quest}~\cite{mot1-meta}, \href{https://www.w3.org/TR/xaur/\%23motion-agnostic-interactions}{W3CXR}~\cite{mot3-w3cxr}, \href{https://learn.microsoft.com/en-us/gaming/accessibility/Xbox-accessibility-guidelines/107}{Xbox}~\cite{mot1-xbox})\\
      & \tabitem \textit{Mot-4.} Ensure controls are as simple as possible, or provide a simpler alternative. (\href{http://gameaccessibilityguidelines.com/ensure-controls-are-as-simple-as-possible-or-provide-a-simpler-alternative}{GAG}~\cite{mot4-gameguide}, \href{https://accessible.games/accessible-player-experiences/access-patterns/do-more-with-less/}{APX}~\cite{mot4-apx}, \href{https://igda-gasig.org/get-involved/sig-initiatives/resources-for-game-developers/sig-guidelines/on-mobility-disabilities/}{IGDA}~\cite{mobility-sig2022guidelines}, \href{\detokenize{https://developer.oculus.com/resources/design-accessible-vr/\%23minimize-the-complexity-of-your-controller-scheme}}{Quest}~\cite{mot4-meta}, \href{https://www.w3.org/TR/xaur/\%23interaction-and-target-customization}{W3CXR}~\cite{vis2-w3cxr}, \href{https://learn.microsoft.com/en-us/gaming/accessibility/Xbox-accessibility-guidelines/107}{Xbox}~\cite{mot1-xbox})\\
      & \tabitem \textit{Mot-5.} Do not make precise timing essential to gameplay--offer alternatives, actions that can be carried out while paused, or a skip mechanism. (\href{http://gameaccessibilityguidelines.com/do-not-make-precise-timing-essential-to-gameplay-offer-alternatives-actions-that-can-be-carried-out-while-paused-or-a-skip-mechanism}{GAG}~\cite{mot5-gameguide}, \href{https://accessible.games/accessible-player-experiences/access-patterns/improved-precision/}{APX}~\cite{mot5-apx}, \href{https://igda-gasig.org/get-involved/sig-initiatives/resources-for-game-developers/sig-guidelines/on-mobility-disabilities/}{IGDA}~\cite{mobility-sig2022guidelines}, \href{https://www.w3.org/TR/xaur/\%23interaction-speed}{W3CXR}~\cite{mot5-w3cxr}, \href{https://learn.microsoft.com/en-us/gaming/accessibility/Xbox-accessibility-guidelines/116}{Xbox}~\cite{mot1-xbox,mot5-xbox})\\
    \midrule
    \multirow{5}{5em}{\textbf{Cognitive \\ Impairments}} 
      & \tabitem \textit{Cog-1.} Avoid flickering images and repetitive patterns. (\href{http://gameaccessibilityguidelines.com/avoid-flickering-images-and-repetitive-patterns}{GAG}~\cite{cog1-gameguide}, \href{https://accessible.games/accessible-player-experiences/access-patterns/clear-channels/}{APX}~\cite{cog1-apx}, \href{https://developer.oculus.com/resources/design-accessible-vr-ui-ux/}{Quest}~\cite{cog1-meta}, \href{https://www.w3.org/TR/xaur/\%23avoiding-sickness-triggers}{W3CXR}~\cite{cog1-w3cxr}, \href{https://learn.microsoft.com/en-us/gaming/accessibility/Xbox-accessibility-guidelines/118}{Xbox}~\cite{cog1-xbox})\\
      & \tabitem \textit{Cog-2.} Ensure no essential information is conveyed by text alone, reinforce with visual and/or speech. (\href{http://gameaccessibilityguidelines.com/ensure-no-essential-information-especially-instructions-is-conveyed-by-text-alone-reinforce-with-visuals-andor-speech}{GAG}~\cite{cog2-gameguide}, \href{https://accessible.games/accessible-player-experiences/access-patterns/second_channel/}{APX}~\cite{vis4-apx}, \href{https://igda-gasig.org/get-involved/sig-initiatives/resources-for-game-developers/sig-guidelines/on-cognitive-disabilities/}{IGDA}~\cite{cognitive-sig2022guidelines}, \href{https://developer.oculus.com/resources/design-accessible-vr-captions/}{Quest}~\cite{vis3-meta}, \href{https://www.w3.org/TR/xaur/\%23critical-messaging-and-alerts}{W3CXR}~\cite{cog3-w3cxr}, \href{https://learn.microsoft.com/en-us/gaming/accessibility/Xbox-accessibility-guidelines/103}{Xbox}~\cite{vis1-xbox})\\
      & \tabitem \textit{Cog-3.} Use symbol-based chat (smileys, etc.). (\href{http://gameaccessibilityguidelines.com/use-symbol-based-chat-smileys-etc}{GAG}~\cite{cog3-gameguide}, \href{https://accessible.games/accessible-player-experiences/access-patterns/flexible-text-entry/}{APX}~\cite{cog3-apx}, \href{https://igda-gasig.org/get-involved/sig-initiatives/resources-for-game-developers/sig-guidelines/on-cognitive-disabilities/}{IGDA}~\cite{cognitive-sig2022guidelines}, \href{https://www.w3.org/TR/xaur/\%23immersive-personalization}{W3CXR}~\cite{cog3-w3cxr}, \href{https://learn.microsoft.com/en-us/gaming/accessibility/Xbox-accessibility-guidelines/120}{Xbox}~\cite{cog3-xbox})\\
      & \tabitem \textit{Cog-4.} Include an option to adjust the game speed. (\href{http://gameaccessibilityguidelines.com/include-an-option-to-adjust-the-game-speed}{GAG}~\cite{cog4-gameguide}, \href{https://accessible.games/accessible-player-experiences/challenge-patterns/slow-it-down/}{APX}~\cite{cog4-apx}, \href{http://igda-gasig.org/get-involved/sig-initiatives/resources-for-game-developers/sig-guidelines/}{IGDA}~\cite{top10-sig2022guidelines}, \href{https://www.w3.org/TR/xaur/\%23interaction-speed}{W3CXR}~\cite{mot5-w3cxr}, \href{https://learn.microsoft.com/en-us/gaming/accessibility/Xbox-accessibility-guidelines/116}{Xbox}~\cite{mot5-xbox})\\
      & \tabitem \textit{Cog-5.} Provide an option to turn off/hide all non-interactive elements. (\href{http://gameaccessibilityguidelines.com/provide-an-option-to-turn-off-hide-background-movement}{GAG}~\cite{cog5-gameguide}, \href{https://accessible.games/accessible-player-experiences/access-patterns/distinguish-this-from-that/}{APX}~\cite{vis1-apx}, \href{http://igda-gasig.org/get-involved/sig-initiatives/resources-for-game-developers/sig-guidelines/}{IGDA}~\cite{top10-sig2022guidelines}, \href{https://www.w3.org/TR/xaur/\%23immersive-personalization}{W3CXR}~\cite{cog3-w3cxr}, \href{https://learn.microsoft.com/en-us/gaming/accessibility/Xbox-accessibility-guidelines/117}{Xbox}~\cite{cog5-xbox})\\
    \midrule
    \multirow{5}{5em}{\textbf{Speech \& Hearing \\ Impairments}}
      & \tabitem \textit{SH-1.} Provide captions or visuals for significant background sounds. (\href{http://gameaccessibilityguidelines.com/provide-captions-or-visuals-for-significant-background-sounds/}{GAG}~\cite{sh1-gameguide}, \href{https://accessible.games/accessible-player-experiences/access-patterns/second_channel/}{APX}~\cite{vis4-apx}, \href{https://igda-gasig.org/get-involved/sig-initiatives/resources-for-game-developers/sig-guidelines/on-auditory-disabilities/}{IGDA}~\cite{auditory-sig2022guidelines}, \href{https://developer.oculus.com/resources/design-accessible-vr-captions/}{Quest}~\cite{vis3-meta}, \href{https://www.w3.org/TR/xaur/\%23spatial-audio-tracks-and-alternatives}{W3CXR}~\cite{sh1-w3cxr}, \href{https://learn.microsoft.com/en-us/gaming/accessibility/Xbox-accessibility-guidelines/104}{Xbox}~\cite{sh1-xbox})\\
      & \tabitem \textit{SH-2.} Support text chat as well as voice for online multiplayer. (\href{http://gameaccessibilityguidelines.com/support-text-chat-as-well-as-voice-for-multiplayer}{GAG}~\cite{sh2-gameguide}, \href{https://accessible.games/accessible-player-experiences/access-patterns/flexible-text-entry/}{APX}~\cite{cog3-apx}, \href{https://igda-gasig.org/get-involved/sig-initiatives/resources-for-game-developers/sig-guidelines/on-auditory-disabilities/}{IGDA}~\cite{auditory-sig2022guidelines}, \href{https://www.w3.org/TR/xaur/\%23various-input-modalities}{W3CXR}~\cite{mot1-w3cxr}, \href{https://learn.microsoft.com/en-us/gaming/accessibility/Xbox-accessibility-guidelines/120}{Xbox}~\cite{cog3-xbox})\\
      & \tabitem \textit{SH-3.} Ensure no essential information is conveyed by sounds alone. (\href{http://gameaccessibilityguidelines.com/ensure-no-essential-information-is-conveyed-by-sounds-alone}{GAG}~\cite{sh3-gameguide}, \href{https://accessible.games/accessible-player-experiences/access-patterns/second_channel/}{APX}~\cite{vis4-apx}, \href{https://igda-gasig.org/get-involved/sig-initiatives/resources-for-game-developers/sig-guidelines/on-auditory-disabilities/}{IGDA}~\cite{auditory-sig2022guidelines}, \href{https://developer.oculus.com/resources/design-accessible-vr-design/}{Quest}~\cite{vis5-meta}, \href{https://www.w3.org/TR/xaur/\%23xr-and-supporting-multimodality}{W3CXR}~\cite{sh3-w3cxr}, \href{https://learn.microsoft.com/en-us/gaming/accessibility/Xbox-accessibility-guidelines/103}{Xbox}~\cite{vis1-xbox})\\
      & \tabitem \textit{SH-4.} Provide separate volume controls or mutes for effects, speech, and background/music. (\href{http://gameaccessibilityguidelines.com/provide-separate-volume-controls-or-mutes-for-effects-speech-and-background-music}{GAG}~\cite{sh4-gameguide}, \href{https://accessible.games/accessible-player-experiences/access-patterns/clear-channels/}{APX}~\cite{cog1-apx}, \href{https://igda-gasig.org/get-involved/sig-initiatives/resources-for-game-developers/sig-guidelines/on-auditory-disabilities/}{IGDA}~\cite{auditory-sig2022guidelines}, \href{https://developer.oculus.com/resources/design-accessible-vr-audio/}{Quest}~\cite{sh4-meta}, \href{https://www.w3.org/TR/xaur/\%23immersive-personalization}{W3CXR}~\cite{cog3-w3cxr}, \href{https://learn.microsoft.com/en-us/gaming/accessibility/Xbox-accessibility-guidelines/105}{Xbox}~\cite{sh4-xbox})\\
      & \tabitem \textit{SH-5.} Allow subtitle/caption presentation to be customized. (\href{http://gameaccessibilityguidelines.com/allow-subtitlecaption-presentation-to-be-customised/}{GAG}~\cite{sh5-gameguide}, \href{https://accessible.games/accessible-player-experiences/access-patterns/clear-text/}{APX}~\cite{vis3-apx}, \href{https://igda-gasig.org/get-involved/sig-initiatives/resources-for-game-developers/sig-guidelines/on-auditory-disabilities/}{IGDA}~\cite{auditory-sig2022guidelines}, \href{https://developer.oculus.com/resources/design-accessible-vr-captions/}{Quest}~\cite{vis3-meta}, \href{https://www.w3.org/TR/xaur/\%23captioning-subtitling-and-text-support-and-customization}{W3CXR}~\cite{sh5-w3cxr})\\  \bottomrule
\end{tabular}
\label{table:guidelines}
\end{table*}
%TC:endignore

\subsection{Analysis}
Interviews were recorded with Zoom and transcribed by a professional online service approved by our University's Institutional Review Board (IRB). Researchers reviewed each transcript to correct errors. We analyzed transcripts using thematic analysis~\cite{braun2012thematic}, starting with two researchers open coding the same set of three participants independently, generating an initial codebook. % We generated an initial list of 1,202 codes, narrowing down to an initial codebook of 829 codes in 172 categories and 21 themes upon their agreement. %\change{Each ``code'' consists of a word or short phrase that briefly summarizes the participants' thoughts on a particular topic. Our ``codebook'' combines these ideas into rough categories and potential themes.} 
One researcher then coded the remaining 22 transcripts according to the codebook. When a new code emerged, the researcher discuss it with the research team and updated the codebook upon agreement. 
%After coding, the researcher condensed codes into a single document. %used Claude 3.5 Sonnet to condense codes into a \change{single} document. %while maintaining association with the appropriate participant. 
%\change{\textit{We ensured participant confidentiality by only sharing researcher-generated codes with the large-language model. No direct transcripts or quotes were uploaded. This method was approved by our University's IRB.} We included our prompt in Appendix~\ref{claude-prompt}.} %\anonymous{We did not specifically mention LLM tools in our consent form, but we contacted our University's IRB office to ensure this analysis method was permitted without listing explicit consent. Since we used LLM tools in a manner more conservatively than someone would use analysis tools like Google Sheets, we did not need to explicitly mention LLMs in our consent form.} 
%The resulting csv file \textbf{generated a list of 3,977 unique codes}, % \change{created by the researcher}
%which we separated into 1,293 ``guidelines'' codes (specific to guideline evaluation) and 2,684 ``main'' codes (for the remaining parts of the transcript).
Researchers then met to derive themes and sub-themes using axial coding and affinity diagramming. %Many themes appeared, particularly those related to common challenges in traditional software engineering roles. \textbf{To stay within reasonable length, we prioritized presenting themes relevant to the intersection of both XR development and a11y, as well as findings that seemed particularly novel compared to prior work.} % to derive a list of six main themes and 26 subthemes.
After identifying an initial set of themes, researchers cross-referenced the original data, the codebook, and the themes to make final adjustments and ensure consistency. Our analysis resulted in a codebook with 3,974 codes, six themes, and 26 subthemes. The theme table with example codes for each theme can be found in Table~\ref{table:themes} in Appendix~\ref{appendix:themes}. 

\section{Findings}
\label{findings}
We present our key findings, revealing the unique XR features that make a11y challenging, developers' practices in XR development, their attitudes towards XR a11y, interpretation and needs on existing XR a11y guidelines, and their preferred tools to assist XR a11y implementation. Our interviews revealed tensions between immersive experiences and accessibility, developers' unique approaches to implementing accessible features, and significant differences in how accessibility was prioritized and implemented across company sizes.
% Numbers of participants listed in this section are those who explicitly mentioned following these practices, though additional developers may follow these practices as well.

% \yuhang{transitions to cover themes. analysis procedure, how many codes, themes, examples}

% \yuhang{Provide codebook in appendix}

% \subsection{Overall experiences}
% [previous section discusses self-taught XR and methods to learn XR]

% ----------------------------------------
%
%          Subsection break :)
%
% ----------------------------------------

\subsection{Why is it Difficult to Integrate A11y in XR?}
% XR + Accessibility Development Challenges
\label{findings-challenges}
Our study reveals unique characteristics of XR applications that bring additional a11y development challenges compared to conventional 2D interfaces, highlighting needs for dedicated considerations when designing XR a11y features. \textbf{Interestingly, these challenges often stem from the very features that make XR compelling}---immersion, spatial interaction, and novel input methods---creating tensions that developers must navigate to create inclusive experiences.

\subsubsection{Sacrificing A11y for Immersion.} 
\label{balancinga11y}
% P3, P4, P5 P6, P10, P11, P13 P18, P19, P20, P22 P23
Nearly half of our participants (12 out of 25) reported a fundamental tension between creating immersive XR experiences and implementing accessibility features. At least three developers (P4, P10, P20) % To address this, we found that developers 
explicitly prioritized their game's \textit{feel} over users' a11y needs. %\yuhang{how many participants?}.
% \textbf{\textit{Prioritizing Game Feel.}}
For example, P4 and P20 hesitated to remove flickering visual effects in their apps although they were aware that such effects may bring risks to people with cognitive disabilities (e.g., triggering a seizure). %---therefore, developers should limit flickering images in their apps. %Although XR simulation apps are common for training or replaying scenarios without real-world harm, 
%However,  %even if they flicker too much for someone with cognitive a11y needs. 
They felt such effects were a necessary component to convey their apps' themes, and they were therefore willing to exclude certain groups of users with disabilities: \textit{``For some app designs, just trying to incorporate certain accessibility guidelines is going to be a \textbf{cursed problem that you cannot solve}. And just to be aware of those problems at a design stage and conscientiously make that trade off, like---we care enough about developing this app that we are willing to exclude this chunk of potential players''} (P4). This perceived tradeoff represents a significant barrier to accessibility adoption in XR, as developers hesitate to implement features they believe might diminish the immersive qualities that define the medium. %\tochange{DK: Could cut down this quote / integrate into our discussion more? --- this quote is interesting, i would keep it}

\subsubsection{A11y Issues and Needs in XR Interactions}
XR applications use unique interactions that enable powerful experiences; however, we find that these same interaction methods create significant a11y barriers that are fundamentally different from those in 2D applications. Below, we detail key a11y challenges specific to different XR interactions, highlighting developers' difficulties in making these interactions accessible. % including XR hardware challenges as they relate to accessibility, as well as how XR developers struggle to effectively accommodate their users. 

% \dk{Moved Depth Cues to Solutions, where it doesn't completely fit either, but maybe does slightly better than here.}
% P3 P23 P25 P12 P4 P16 P22 P20 P24 P9 %(p21 text can be ambiguous)
\textbf{\textit{Text Resolution Harms Legibility.}}
Ten participants mentioned concerns that reading text on XR platforms was a major challenge that caused a11y barriers due to low screen legibility, especially for people with visual impairments. %\yuhang{update}. 
Notably, even for users without visual impairments, factors like display resolution and small screen sizes can make text-based communication difficult to interpret: \textit{``One day I can get the headset to sit on my head just right and all the text is clear and I can read things and its great. But then the next day for whatever reason, it's just is not sitting just right and everything's blurry... I have 20-20 vision. So if that's a problem for me, it's a problem for everyone else''} (P4).
Eight developers % P1, P2, P9 P12, P16, P20 P22, P23
limited the use of text as much as possible and instead prioritized symbols, graphics, and/or large virtual objects to convey information. However, five participants (e.g., P6, P17) highlighted that making graphics accessible brings more complex technical issues, such as how to add alt text to 3D objects and how to develop a platform-supported 3D screen reader to read the descriptions---neither of which have standards in XR. %(P6, P8, P17, P18, P24)
% Although HMD screen technology has improved in recent years, older hardware (e.g., the Oculus Rift CV1 VR headset) or small displays like Brilliant Labs' Frames smartglasses (only 5.8mm diagonal, 640x400, 20\textdegree FOV) limits readability, necessitating new a11y solutions that don’t occur in 2D interfaces. 

Some big tech developers P23 and P25 believe XR screens are in an ``unacceptable'' (P23) middle ground, where text illegibility greatly detracts from immersion (P25)---users are aware what they are looking at is text, but they spend a long time trying to read it. P23 highlighted this issue: \textit{``When things are in the middle, people spend more time, they spend more strain, they're more distracted as a more taxing task.''} XR text's illegibility also poses barriers to some potential a11y features. For example, P25 developed a head-worn AR transcription and translation feature, but users had to perpetually concentrate on the device's small screen instead of their outside environment, missing details when another person is speaking. This missing context is particularly problematic if the AR user %has an impairment that 
needs both transcription and lip-reading to understand others. 
\textit{``It's really hard for them to [focus] on other people, right, other activities, because you can only focus on one thing and do it right. ... And it's also very tiring, because you needed to fully focus on it for quite a long time''} (P25).

% Many MR and VR platforms do enable users to physically walk closer to world anchored text to read signs more effectively (P19), but this solution does not aid users with motor impairments, %\yuhang{any participants mentioned this?}, 
% \dk{no participants mentioned challenges in this paragraph, can cut}
% text on signs outside of walkable areas, nor if the users doesn't know the sign's language. % \yuhang{what does this mean?}. %Furthermore, text rendered as into 3D often becomes generic objects,  potentially losing metadata that labels them as text.
% if for disability, e.g., low vision, 3D text for low vision, motor disability; limiting the use of text in XR, which makes accessibility harder for screen readers describing shapes/graphics

\textbf{\textit{Spatial Audio Opportunities and Challenges.}}% \kz{rephrase this as a challenge / need to better echo the themeL A11y issues and needs in XR}
\label{spatial-audio-challenges}
%Developers P18 and P19 believed that spatial audio afforded by XR headsets is a key to making apps more immersive. 
While spatial audio creates new accessibility opportunities by providing orientation cues and information about the virtual environment for PVI, %, applying traditional 2D solutions like captions requires careful reconsideration. More broadly, we find that this belief creates a unique opportunity for a11y, as spatial audio can provide orientation cues and information about the virtual environment. 
% However, users with one ear, for example, require additional directional cues than an existing 2D accessibility solution like mono audio can provide.
% Accessibility solutions targeting speech \& hearing impairments in XR 
%\yuhang{in XR? because they are spatial audio? need to say that explicitly}
% extend past 2D solutions (e.g., mono audio), instead needing more present, visual cues to grab user attention and convey directional audio. 
% P18 argues that by using headsets, apps are much more immersive, and spatial audio is a large factor in bringing people in.  %Therefore, mono audio no longer becomes an accessibility solution \yuhang{i think you are skipping a lot of logic here, why mono audio was an accessibility solution? for whom?} 
% if spatial audio is a core part of the experience: \textit{``A lot of [our] attention-grabbing work is by spatial audio. [...] Usually, we will design a sound effects system in the beginning, [...] even before we established the whole thing. We will try to do some sound effect to simulate the kind of experience we want to do''} (P18). \yuhang{again, where is the a11y issue from the quote?}
applying 2D a11y solutions for d/Deaf and hard of hearing people, such as closed caption and subtitle systems, requires thoughtful reconsideration for XR. %must be thoughtfully reconsidered for XR. %\yuhang{why did we suddenly jump to closed caption and subtitle? are there current solutions in XR?}
P19 suggested developers spatially-anchor captions to audio sources, displaying a caption in view when nearby (e.g., ``people chatting''); but others mentioned 
%However, current caption systems in XR do not innately indicate sound direction, nor do heads-up displays (commonly found in 3D titles) work well for VR apps.
%however, participants disparaged 
attaching anything to users' heads as obtrusive and hard to focus on (P6, P10): \textit{``[It's like] putting a sticky note on somebody's face''} (P10).

\textit{\textbf{Movement and Interaction Barriers.}} Several participants (P1, P13, P16) noted that common XR navigation methods like teleportation, while beneficial for some users with mobility impairments, could create disorientation for users with visual impairments who need to reorient themselves after teleporting. Additionally, the precision hand movements required by many XR apps present barriers for users who cannot operate standard controllers (P13, P14).

 %(P2, P11, P13, P14, P16).

% Furthermore, interacting with XR objects typically shows a corresponding response on the user's device, but developers expressed concerns that these interactions may not provide the user with the expected tactile feedback, widening a cognitive disconnect between what the user sees and what they feel. These concerns are particularly prevalent in AR and MR platforms, where users commonly interact through a smooth touchscreen or in the air without controllers. P18 for example needed to consider the types of interactions they could do that wouldn't feel ``awkward'' in their AR apps, deciding to focus on conceptually virtual objects (e.g., magic spells) for users to stay immersed without touching them directly. \textit{``You know, pushing a ball ... if you render it in [a] really realistic setting, people will feel like it's supposed to be touchable, but you actually cannot touch it. Right? That's a contradiction in your mind''} (P18).

\subsubsection{Limited Representation of Diverse Bodies}
% \textbf{\textit{Normative View of an XR ``User.''}}
% \yuhang{what is the overall finding, e.g., XR is missing inclusive representation for people with disabilities? You start with example directly.} 
Developers expressed concerns that XR systems inadequately represent users, creating barriers to inclusion. For example, VR headsets (e.g., Meta Quest 3) overlay holographic hands tracking the user's real-world hand position, allowing users to see their hands in front of virtual objects. However, these hands show a ``typical'' representation of a hand, and headsets have difficulty tracking users' missing limbs or fingers (P14, P19). As P14 explained, \textit{``[VR] has this normative view of what would be a hand, how do you interact with your hand, whereas not necessarily all people have the same level of ability in the hands or other body parts...The design of headsets and controllers---they're specifically made with some kind of user in their mind''}. This finding aligns with prior work on avatar representation for PWD~\cite{gualano2023invisible,mack2023towards,zhang2022its}.%, and these works should extend %from external, social VR applications 
%into self-viewable, OS-level interfaces. For example, VR headsets (e.g., Meta Quest 3) overlay holographic hands tracking the user's real-world hand position, allowing users to see their hands in front of virtual objects. %, since their real hands would otherwise be occluded by the virtual elements. 
%However, these hands show a ``typical'' representation of a hand, and headsets have difficulty tracking users' missing limbs or fingers (P14, P19). %In other words, if users are missing limbs, they may not be able to use apps, e.g., those that require use of two hands. For users with two hands but are missing one or more fingers, then headsets may have trouble tracking and the user's virtual hand is misrepresented.

% Developers like P14 also question the ability to use alternate control methods to accommodate users with difficulties using default methods: \textit{``The design of headsets and controllers---they're specifically made with some kind of user in their mind''} (P14).

\subsubsection{Performance Constraints of XR Platforms Limit A11y}
\label{performance}
% \yuhang{first sentence should be key point, not background.}
% P1 2 6 8 9 19 20 21 22 24 25
% a11y: 8 19 21 24 25 
Eleven participants expressed practical concerns about the performance requirements of their XR applications, particularly as many users move from PC-based VR systems to standalone, mobile devices with lower computational power. %This trend greatly lowers performance ceilings of applications. 
%With the limited computational power, 
Five participants noted that adding logic for PWD-exclusive a11y features would affect performance overhead (e.g., P19, P21, P24), especially for the XR apps that need high-quality visual effects (P11).
%Some developers have mostly adapted to mobile performance limitations (e.g., P1)%\textit{``I've gotten so used to working with a very limited number of polygons and a very limited number of draw calls.''}
%, but performance limitations still affect P11, whose startup's high-quality visual effects require a computer-bound headset (e.g., Rift CV1). 
Paradoxically, performance issues may itself introduce additional a11y problems, like flickering images that may trigger photosensitive people (P21). %Furthermore, many current VR a11y features require an external device (e.g. keyboard (P19) or non-standard controllers (P14)) which are difficult to apply without connecting to a PC.
% (e.g. Canetroller~\cite{zhao2018canetroller})
% or require additional setup when connected to a PC. %future support methods need to be contained in a standalone device.

% \yuhang{the evidence is not relevant to a11y...you key findings need support from the participants, but the participant only said that the performance is limited, and did not say that this affect their a11y implementation.}

\subsubsection{Challenges with Onboarding First-Time Users}
\label{onboarding}
% A11y solutions benefit everyone, but similarly, some challenges using XR platforms harm everyone--- even people without accessibility needs--- which may create further a11y challenges for PWD.

% P1 P12 P21 P22 P18
% \textbf{\textit{First-time User Unfamiliarity.}}
As many XR interaction techniques are unique from users' prior experience, XR platforms often require learning curves. Five developers considered first-time VR users as people with accessibility needs, %particularly given that unfamiliarity may cause users to not recognize objects as interactive or become confused when trying to interact with static objects (e.g., P12). 
as novel interaction techniques require careful explanation and practice.
While video tutorials may help developers learn XR tools, users need a hands-on tutorial, % to teach them each interaction and correlate them to a specific action (P18)---
which may not be feasible after app distribution (P18). %As P18 noted, \change{\textit{``People learn [interactions] corresponding to [something they're familiar with]. So that's [a] very important part%like so that we also do some tutorial doing that... Even a video tutorial doesn't work.''}} \tochange{DK: Quote is a bit hard to read; should we cut down or reshuffle? --- can remove; but then need to add some participant numbers to prior sentence.}

\subsection{How do Current XR Development Practices Impact A11y Integration?}
% Development practices
\label{findings-practices}
Prior work has identified barriers to accessibility integration in traditional software, but our findings reveal how XR-specific development practices create unique implementation challenges. We found distinct patterns in how XR developers approach their work that significantly influence their ability and willingness to incorporate accessibility features.
% Next, we present participants' current development practices, highlighting specific XR practices that significantly affect developers' ability to integrate a11y features into their preferred platforms, thus informing recommendations for improving XR accessibility support.
% Next, we present participants' current development practices, identifying several XR development practices that significantly impact developers' ability to implement a11y features in their preferred platforms, to better inform how XR accessibility support methods should be produced.%, including general practices for XR development in their preferred platforms and what barriers they face, to better inform how XR a11y support methods should be produced. %\tochange{R2 asked about accessiblity maintainess process, might be able to add a couple of sentences in this section? " Limited exploration of how accessibility features are maintained through updates and fixes" - R2}
%\yuhang{clarify why the practice is important to our research question---XR A11y?}
%\tochange{DK: A11y maintenance is explored in Discussion instead of here \S5.2.3}

% ----------------------------------------
\subsubsection{Knowledge and Role Distribution in XR Teams}
\label{lack-training}
%Participants lamented having a sporadic knowledge base towards XR development, making it difficult to know what plugins are available (P1). 
% \dk{need a more compelling transition here}
Supporting Wang et al.'s findings~\cite{wang2025understanding}, we %XR developers commonly learn XR development practices outside of conventional settings. We 
found most of our participants (17 out of 25) had little to no formal XR training, instead learning through online resources (14 participants), % 1, 5, 6, 7, 8, 11, 12, 13, 16, 17, 19, 20, 22, 24
% like courses (P5, P13, P20, P22), tutorials (P5, P11, P12), forums (P6, P22, P24); 
on the job (8), %(P1, P7, P8, P11, P12, P17, P19, P25).
or from personal projects (3). %(P6, P11, P14).
%Developers read books (P20) or used platforms like YouTube (P1, P11, P16, P20, P22), Udemy (P5, P13), and/or Udacity (P5) to teach themselves best practices. Developers consult online documentation ($n=8$) and forums for assistance, including official community forums ($n=6$) and unofficial Reddit ($n=2$) and Discord servers ($n=3$). 
Furthermore, zero participants indicated formal a11y training. Participants lamented this self-taught experience left significant knowledge gaps for a11y implementation regarding accessibility guidelines, tools, terminology, and best practices. %Six participants %(P1, P7, P14, P15, P17, P21)
%specifically mentioned lacking knowledge on guidelines, existing tools, best practices, and appropriate terminology to use when discussing a11y. 
% Participants lamented this self-taught experience left significant knowledge gaps for a11y implementation.%, which proves challenging when finding best practices to implement a feature. %For example, P1 explained their sporadic education as \textit{``just kind of a base of knowledge that doesn't get filled in an order. [It's] fun, but there's definitely blind spots in my development skills''}.
%Formal training for XR A11y was even more rare; 

% For example, even as an educator, P2 is \textit{``not an instinct in any VR class [she's] developing''}, and does not intend to include accessible design practices into her teaching. By contrast, P4, who frequently incorporates a11y starting from the design stage, cited \textit{``the problem of accessibility in VR [as] more just ignorance of the problems''}, emphasizing a need for popular apps' accessibility features to be better highlighted.

% \subsubsection{Diverse Responsibilities within XR Teams.}
\label{responsibilities}
Although all 25 participants considered themselves `developers,' we found that their work included many responsibilities beyond programming. Unlike traditional software development where developers commonly have one specialized role, our participants juggled multiple simultaneous aspects of development, including programming, %(e.g., software developer), %, interaction developer, user interface developer)
to management, %(e.g., CEO), %creative director, manager, CEO) 
design, %(e.g., asset designer), %, level designer, UI designer
%writing, research, 
research, and animation. %\textit{Quantity} of job responsibilities varied slightly by company size, with freelance and startup employees often spread out over many roles, while midsize and big tech employees typically had one programming-related role and up to two secondary roles (e.g., software engineer, interaction designer, and team manager).
These diffusion of responsibilities leads to unclear responsibility assignment for a11y tasks, restricting developers' ability and willingness to effectively learn and implement a11y features. We found that company size influenced role specialization, with larger organizations allowing more focused responsibilities but also creating coordination challenges for accessibility implementation, while  freelancers and startup employees were particularly stretched across responsibilities. We have included developers' primary roles alongside their years of experience in Table~\ref{table:participants}. %as they're responsible for finishing multiple aspects of an app by a limited deadline.

\subsubsection{Dynamic Code Generation Creating Technical Barriers}
%Researchers have been investigating how to enable a11y labels to virtual environments to support features like screen readers~\cite{zhao2019seeingvr}; therefore, we further analyzed developers' practices in code generation to understand the feasibility of different 3rd party a11y support tools in XR. %To mitigate data loss associated with multiple developers pushing changes to Unity prefabs (drag-and-droppable object templates in Unity) or Unreal Blueprints, 
To enable easier version control and code merging, we found that multiple big tech developers (P3, P15, P25) prefer to dynamically generate their scenes and virtual objects through code instead of visual editors like Unity prefabs (drag-and-droppable object templates) or Unreal Blueprints. As P3 explained, \textit{``It's far easier to deal with merge conflicts when you can see the code that's building out your scene, right? ... When working on HoloLens apps in Unity, [if] someone changed the prefab, now my day's work is hosed. Because the stuff I did in the GUI with clicking buttons in 12 different places is gone.''}
While this approach improves collaboration, it creates challenges for implementing third-party accessibility tools that often integrate through visual editors \cite{zhao2019seeingvr}. %Big tech developers including P15 and P25 keep relatively clear scene graphs, dynamically generating their scenes through code instead of an abundance of prefabs. When prefabs and scene changes are necessary, developers work in their own copies added to their .gitignore. 

\subsubsection{Reusing Code Between Projects} 
\label{reusing}
Developers have a variety of considerations when starting new projects, including what packages to install, whether they start from a template or an empty scene, and what packages to include. These findings help inform a11y support methods to ensure compatibility and reveal how willing developers are to start from an ``accessible'' design template or incorporate a11y packages into their workflow.
% \yuhang{need transition sentences to clarify why we care about these in XR A11y contexts. And for each subtheme, need to connect back to XR A11y} 
% Due to constantly shifting guidelines, 
Most participants start new projects from scratch (7 developers), %(P2, P5, P10, P11, P13, P18, P19, P25)
with P10 arguing against the use of templates to keep her projects lean and reduce per-project development time by only adding components she needs. Five participants  % (e.g., P11, P12, P14, P20, and P25) 
commonly start from template scenes that include existing prefabs or packages that they customize to fit their needs, with P12's startup having a premade, base template scene they base all projects on. %Given P12's company's resistance to a11y implementation, they may be less likely to include accessible prefabs in their base template without thorough investigation. 
Although participants varied in the exact prefabs they used when starting their projects, most participants started with a player component (P1, P5, P8, P13) %P1, P5, P8, P13
and an XR camera and tracking prefab (e.g., Unity XR Interaction Toolkit's XRRig, 7 participants).% or Meta XR Core SDK's OVRCameraRig (P8)). % 1 5 10 12 13 21 22

However, even developers who start from scratch reuse code between projects (7 developers), particularly existing components like a customized player class. For freelancers specializing in a particular area, like P1 making custom VR video applications, they often use the same base scripts with slight customization based on preferred input methods for their target platform, writing reusable components to save them time in the future (P6). Some big tech companies maintain an internal knowledge base for employees to reference (P25)---their maturity enables them to move more quickly on certain 2D a11y features like screen reading or alternate language fields by applying reusable templates and technologies (P23). However, such a11y feature templates do not yet exist for XR development (e.g., P6, P15, P22).

\subsubsection{Limited A11y Testing with PWD} %We report developers' practices and challenges with a11y-related app testing.
\label{testingpwd}

%\textbf{\textit{Lack of A11y-Focused Feedback for XR.}}
Multiple developers noted user feedback as vital to informing new features (e.g., P2, P23, P25), but we found a significant gap in developers' testing practices with PWD, opting instead for a general sample of users (6 participants) or no user feedback at all (P19).  %Developers only perform internal testing by way of focus groups/demos () %P23, or test with a general sample of users rather than an a11y-targeted subsample (). %P17 P4 P8 P18
% Some startups lack any user feedback at all (P19), but those that do test outside their organization lack PWD-targeted testing groups, opting for a general sample instead %(P2, P4, P18, P21)(5 participants. 
General samples do produce some testers with disabilities, as P2 explained: \textit{``If [you] have 100 people try something you're usually going to see...some wheelchairs, you're gonna have small children. And you're going to have people with limited usage of both hands, right. So those sort of things sort of naturally come up,''} (P2). However, this approach misses many a11y needs and provides limited feedback for improving a11y features. %but developers do not seek PWD-specific testing groups for their apps. %Big tech companies that \textit{do} provide resources to perform a11y-specific testing often move too slowly for the feedback to be useful (P3): \textit{``Every time we've tried to do user studies, unrelated to mixed reality, it's like, two months later, we get the data back, and [by then] the entire feature's changed ... and is not useful anymore. So I think it's just the speed of a big company.''}

\subsection{What Motivates and Hinders XR Developers to Integrate A11y?} 
% Perceptions and Attitudes Towards XR A11y Integration
\label{findings-attitudes}

% We found developers 
% We now outline the attitudes that developers have towards implementing accessibility features into their XR apps, including their motivations at various company sizes.

% ----------------------------------------

% \subsubsection{Diverse Attitudes Across Organizations}
\label{motivations}
%\tochange{if you have time, try to connect to the definition of disability across this section; did developers conceptualize disability as personal traits, so that they don't think XR is for certain disabilities? Or any of them follow the social model?---so that this is not only discussed in Discussion, we can sort of pave it in findings too.} %DK addressing at the end of this section
Prior work has examined general barriers to a11y implementation, but our research reveals how organizational contexts fundamentally shape developers' motivations and constraints regarding a11y. We not only asked developers to describe a11y features they've already made, but also what would encourage them to implement new a11y features in their apps. 
Our developers expressed concerns that the overall financial \textbf{cost} of implementing a11y features was \textbf{``worth it''} to their organization (15 participants). %p16 7 1 13 4 23 8 9 12 22 25 19 20 11 15
As such, they would primarily be motivated by knowing that a large \textbf{number of users} with disabilities would use the features addressed (21). %2 P3, P4 P7, P8, P9, P10, P11 P12, P13, P14, P15, P16, P17, P18, P19, P20, P21, P22, 24 P25
Developers seek hard data on how many users they would gain or benefit by addressing each feature, as well as how many users they \textit{could} reach via positive \textbf{reviews} on online marketplaces (P1, P17). % 1 17
Developers would also benefit by knowing the estimated \textbf{time} necessary to implement features (19), % 1 2 16 14 12 13 4 5 6 7 8 9 10 18 19 20 23 25 15
or an approximate metric of difficulty vs. value of each system (P6, P12, P19). Finally, some companies face \textbf{legal} concerns by not incorporating accessibility into their apps, leading to a11y feature additions not by developer choice (6). %P3, 12 15 17 19 P22.
% We outline each of these motivations as follows, distinguished by the unique differences of how they affect developers at companies of each size, from freelancers to big tech:
We outline each of these motivations below, finding distinct patterns in their effects across different organizational contexts:

\subsubsection{Freelancers: Client-driven A11y} Freelancers reported mixed attitudes towards implementing a11y features, %but are generally more apt to do so than startup developers. %(when not building a tech demo\footnote{P6's project shown during the interview was intended to help them self-teach VR development, citing their app \textit{``[Wasn't] made for a super mass consumption, right? It was like, maybe some other people play this, I just need to make it exist and work first, accessibility is definitely a second priority for me.''}}). 
%However, spending time implementing a11y features is 
largely dependent on their clients' wishes. Their ability to include a11y features depend on whether those features serve the client's target audience (5 participants). %P1, P2, P5, P20, P21). 
%These clients also only care about final built deliverables rather than intermediary steps, so freelancers are incentivized to deliver a project to-spec by a specified deadline with no additional features. 
%P4 contrasts their a11y priorities when working on freelance projects versus projects they lead: \textit{``For a project I'm leading, I would expect 25, 30\% [of my time towards a11y development]. But for a project [where] I'm just a coding monkey, ... I would make my opinions on accessibility best practices known, but I would not die on that hill. I would make suggestions, but whatever the client wants is what gets me paid''} (P4).
% When provided the clearance to develop accessibility features, freelancers are more or less motivated dependent on their knowledge and exposure to the development platform, as well as the target audience of their app. but if they can't convince their clients to pick a more accessible design or the features are outside the scope of the target audience, then those features will get dropped. 
For example, P1 created a video player for a church pastor, who's primary audience was older, first time VR users. Therefore, they preferred a more simple UI as \textit{``more usable to the target audience.''} (P1)

\subsubsection{Startups: Prioritizing Survival Over A11y}
Startups demonstrated the most reluctance toward implementing accessibility features, with five participants describing their companies as being in  ``survival mode,'' not targeting PWD. We found that startup developers cannot directly convince their clients of accessibility needs like freelancers can, instead reporting to a project manager who assigns goals (5 participants). % 22 25 15 24 16
These developers not only questioned the number of users with disabilities who would benefit from accessibility features but also used dismissive language regarding potential users. As P13 stated: \textit{``How many of them \textbf{want} to engage in VR games? ... [It's] not fair to expect somebody to literally take like three years of their life and make it completely not valuable to them in any way just to allow the access of \textbf{these other people}.''} This perspective reflects the intense market pressure startups face and their tendency to prioritize features that reach the largest user base.
% Other developers talked down about startups' lack of a11y implementation, referencing startups' reliance on good reviews as a motivator: \textit{``In a sense, accessibility won't be done until someone complains about it, I mean, sues or just can't use it''} (P3).
% P13 in particular also lacks realistic perceptions of how long it takes to develop accessibility features, with P13 stating that \textit{``It's just not fair to like expect somebody to, to literally take like three years of their life and make it completely not valuable to them in any way just to allow the access of these other people, you know.''}

\subsubsection{Midsize: Emerging A11y Infrastructure} Companies like XR engines, universities, and medical facilities occupy a middle ground with emerging but limited accessibility teams. Midsize companies may not have the same resources as big tech, but they also do not need the same ``survival mode'' pressure as startups since they've already established themselves.  %don't have the same high-profile legal teams as big tech companies, but also don't have to worry about poor user reviews as startups do since they've already established themselves. 
% P17 was notably the only employee with an accessibility team at a midsize company, and the size of their team at the time of our interview was only two people.
P17, who worked at a midsize company, noted their accessibility team consisted of only two people.%and the focus of their work was primarily web development, not XR
In an effort to encourage developers to take more agency, P17's team takes an interesting approach, reassigning a11y bugs back to developers that sent them out: \textit{``So they took ownership of the accessibility changes that they need to make, even if ended up being more effort in the long run to get them to do that''} (P17). This strategy demonstrates how midsize companies attempt to balance a11y needs with limited resources.

\subsubsection{Big tech: Compliance-Driven A11y}
Surprisingly, while big tech companies demonstrated the most robust approach to a11y with dedicated teams and established standards for 2D applications, these practices did not fully extend to their XR development. %Developers at big tech companies appear to care about a11y the most, with full dedicated a11y teams in their organization to ensure compliance with base standards. Although big tech companies have relatively mature a11y teams and standards for \textbf{\textit{2D}} development, similar practices do not fully extend to XR platforms. 
%Despite existence of a11y teams, they largely exist outside XR development teams, therefore attempting to apply 2D or 3D standards to XR apps which do not fully enable accessibility on these platforms. 
While big tech has internal, automated tooling for a11y compliance, they appear rather surface-level, such as checking if a button performs its expected action when pressed. But these tools lack proper user evaluation for scenarios unique to XR, where a button is out of reach or the user can't see the button due to objects in front of it (P21).
% these tools merely scan objects like flat panels for their isolated a11y compliance---not the scene as a whole.
P3 indicated that big tech companies that \textit{do} perform user evaluations often move too slowly for the feedback to be useful: \textit{``Every time we've tried to do user studies, unrelated to mixed reality, it's like, two months later, we get the data back, and [by then] the entire feature's changed ... and is not useful anymore. So I think it's just the speed of a big company.''} P3 also simplified the rationale for caring about a11y to be a fear of being sued, with big tech a11y teams sticking to a bare minimum to ensure compliance.

\subsection{How do XR Developers Implement A11y Solutions?}
% Developers’ Current A11y Solutions
\label{findings-solutions}
Despite limited resources and guidance, we found that many XR developers have created innovative accessibility solutions that could inform future standardization efforts. Participants described various approaches they've successfully implemented, highlighting practical strategies that could be more broadly adopted.
% Developers have already implemented a variety of accessibility features in past projects, offering the following as implementable with some considerations:

% ----------------------------------------
\subsubsection{Alternative Input Methods}
\label{io}
Developers have integrated diverse input mechanisms to accommodate different accessibility needs. 
% Developers use a variety of custom input and output methods to enable accessible features in their XR apps to make up for lacking sensors or inaccessible controllers.
Participants reported using controllers like the Xbox Kinect (P3), Nintendo Switch JoyCon (P3), the Xbox Adaptive Controller (P20), and the Merge Cube (P4) to enable more accessible interactions. To address compatibility challenges across platforms, %alternate input methods in their XR apps for users with motor impairments. Use of alternate input methods introduce challenges with standardizing input across devices, which some developers mitigated using toolkits 
some developers used libraries like Rewired~\cite{rewired} (P5) and Unity's XR Input system. In an effort to make their platform more accessible to alternate input methods, P17's team incorporated keyboard navigation into their platform's controller system, \textit{``So any game that already had controller support, now suddenly had keyboard support. So this is a big feature for instantly opening up keyboard control to, you know, 1000s, millions of games that already had controller support.''} This platform-level solution demonstrates how accessibility features can be efficiently scaled.

\subsubsection{Leveraging Platform A11y Features.} 
\label{nativedevicehooks}
% As opposed to ``high-fidelity'' XR development platforms (e.g., Unity)~\cite{nebeling2018trouble}, some developers noted advantages with ``Native'' frameworks (e.g., Babylon Native, P3) as able to convert programs to OS-native commands, thus enabling the a11y supports (e.g. VoiceOver and Android Talkback) on mobile device. %In contrast, P3 complaint that Unity lacks the ``hook'' into OS-level equivalents (e.g. VoiceOver and Android Talkback) compared to Unity, which uses a \textit{``weird mashup of shapes and random prefabs''} (P3) to build user interfaces. 
%P17 echoed this sentiment, mentioning that \textit{``Neither does the Roblox engine, it's, you know, entirely just 3D rendering. There's no hooks for these for [a] screen reader.''} 
Some developers strategically used their platform's capabilities to implement a11y features. For example, P3 created an accessible mobile AR application by %used Babylon Native to create a visually impaired accessible, mobile MR measurement and 3D model visualization tool. The app enabled users to measure distances in a room by moving their phone outward, but it was also fully compatible with the platform's native screen reader by 
adding invisible 2D buttons atop 3D objects so their mobile device's screenreader could read them: \textit{``These invisible buttons, you could cycle through just as you would any other button. And they could kind of highlight with the screen reader focus indicator, ... And it would ... give you an object, or it'll basically tell you the manipulation of the 3D objects that you would do. And it was directly in line with like that other buttons on the screen. And so it kind of was, hopefully, it was an intuitive way to basically have buttoned controls to move 3D objects in space.''} This approach demonstrates developers' ability to leverage existing platform capabilities rather than developing new a11y tools from scratch. %\yuhang{describe the method of converting object to button method by P3}

\subsubsection{Examples of Existing A11y Solutions}
Our participants have successfully implemented many common a11y features for their XR apps:

% Colorblindness (7), one-handed mode (5), and closed captions/subtitles (6) %12 19 5 22 18 3
% were among the most common existing a11y features developers implement in their projects. This section outlines accessible XR solutions developed to address particular conditions, sorted by disability type mirroring our guidelines.

\textbf{\textit{Solutions to Address Visual A11y.}}
\textit{Colorblindness} was the most common impairment addressed by developers (7 participants), with three participants (P4, P20, P23) using both shape and color to demonstrate essential objects. P15 and P23 recommended implementing colorblind-accessible color palettes, offering systems that change colors application-wide on the fly, similar to how Beat Saber~\cite{beatsaber} supports custom color options (Appendix~\ref{appendix:beatsaber}) %\tochange{add the figure to appendix and refer to it}. %Beat Saber, shown in image (A) of , now supports custom color options for a variety of objects, benefitting colorblind players and those seeking personal style. 
Other developers implemented adjustable text size (P2, P22), legible fonts (P9), and magnification (P2, P3, P22) features. %, and window resizing (P2).
P3 notably implemented a custom AR screenreader using a system of on-screen invisible buttons. 
Developers also used depth cues in VR, including non-interactive elements like fog, to provide a sense of 3D space (P1, P15) for people lacking depth perception. In this regard, we also found that some VR hardware affordances innately provide disability \textit{solutions}, as stereoscopic HMDs have enabled users without depth perception to experience depth for the first time: \textit{``[My user] was able to have a sense of depth through this VR experience. And for the first time, he said forever in his life''} (P1).
% However, we found that developers are confused what constitutes non-interactive elements in VR. When considering whether to implement an option to hide non-interactive elements, P1 questioned whether users' mere presence in the scene constitutes interaction, saying: \textit{``[Is] directional light interactive or not while it's casting shadows? And [if] I'm interacting in the shadow casting[,] is that interaction or is that distraction?''}. 
% (\[Figure: screenshot from Beat Saber's colorblind settings\]). 
%3D games like Overwatch 2~\cite{overwatch} take colorblind support one step further, providing presets for a variety of conditions (Protanopia, Tritanopia, Deuteranopia) for an existing palette of Primary, Secondary, and two accent colors.

\textbf{\textit{Solutions to Address Motor A11y.}}
% P1, P2, P4, P23, P24
Five participants developed one-handed interaction modes, enabling full functionality with only one controller. %We found that one-handed modes were useful for people with permanent and situational a11y needs---to make up for situational one-handedness~\cite{microsoftInclusive101} in manufacturing, 
To address movement restrictions, P23 notably incorporated their HoloLens' eyetracking capabilities to enable remote selections, while
% Alternatively, P2 had notably not considered to implement a one-handed mode until a tester in her general testing group only had one hand.
% Additionally, although locomotion in XR is a continually researched problem~\cite{DiLuca2021Locomotion}, %P1 highlighted their app's limited movement mode, and 
P4 notably avoided the problem entirely by making their app stationary---\textit{``so we don't need to worry about smooth versus teleportation locomotion, or people being able to physically get up and ambulate around. So someone in a wheelchair can use this [app] just as well as anybody else.''}

\textbf{\textit{Solutions to Address Cognitive and Hearing A11y.}}
% We found that some developers implemented a variety of a11y suggestions, many of which could be considered good general accessible design practices. Some 
Participants incorporated conventional cognitive a11y solutions, including limited animation (P1), simple UI/UX (P1, P18), and pausing (P4). %button to halt their app's execution. P1 also emphasized the use of simple UI design to aid first time VR users, and P18 echoed this sentiment with simple UX interactions in their MR apps. 
%\dk{next part feels a little out of place but a lot of developers mentioned it and it fits here the best of other a11y categories:}
%Eight participants %(P1 P3 P9 P12 P19 P23 P24 P25)
%also mentioned \textit{localization} as a priority, incorporating iconography into their app to minimize the amount of text to translate. Despite not prioritizing localization, P2 and P25 both incorporated automatic translation to their apps. 
% \textbf{\textit{Solutions to Address Hearing A11y.}}
To address hearing difficulties in real-life and virtual environments, six participants % P3 P5 P22 18 25 12
developed subtitles or closed captions for their XR applications, with P25 developing AR and VR transcription applications that reads out audio from sources around the user. %However, P12 would like better guidance on scenarios to use subtitles instead of closed captions and vise-versa, with P3 believing full closed captions are too intrusive in some cases. 
%Additionally, VR headsets like the Meta Quest have downward-firing speakers located near users' ears but do not fully encapsulate them. Therefore, 
P1 also expressed concerns with being able to hear their application at an adequate volume in a busy hospital environment, which they solved by pitching up their narration to boost their app's volume.

\subsection{How Applicable are 3D Guidelines to XR A11y?}
% Evaluating Guidelines for 3D Virtual Worlds
\label{findings-guidelines}
While a11y guidelines exist for traditional platforms and some 3D environments, our findings reveal significant gaps in their applicability to XR's unique interaction paradigms. 
Developers found existing guidelines unable to be sufficiently applied to XR. 
%We further report developers' understandings and implementation suggestions for the a11y guidelines for 3D virtual worlds, investigating the applicability of these guidelines to XR.
% As mentioned in the previous section, perhaps the most prevalent concern from developers throughout guideline evaluation was how many users would benefit from these features being developed.
% We outline the following key findings from developers' feedback on accessibility guidelines for 3D virtual worlds:
% ----------------------------------------
\subsubsection{Limitations of Current Guidelines Applied to XR}
Developers were familiar with some existing a11y guidelines, but each lacked knowledge of how to apply them to XR apps. For example, Microsoft's MR Accessibility Standards~\cite{MSFTa11yMRTK3} are not considered ``fleshed out'' by P3 nor P20, and the W3C's Web Content Accessibility Guidelines (WCAG)~\cite{w3web} %that developers are familiar with 
only apply to 2D interfaces (P3, P17). XR remains a lot less mature than web interfaces (P24), lacking simple system-wide a11y features like zoom despite prior work demonstrating their feasibility (e.g., SeeingVR's magnifier tool~\cite{zhao2019seeingvr}).

When evaluating our presented guidelines, developers generally reported understanding most guidelines (e.g., Vis-1, Cog-2, SH-3) and correlated their understanding to prior solutions they've seen or personally developed in and outside of XR (e.g., referencing Fortnite's visual radial sound indicator~\cite{delaney2021fortnite} satisfying SH-1; P17, P18). However, some guidelines are too broad or use ambiguous language, e.g. when to use text, symbols, icons; or subtitles versus closed captions, and how each should appear in their XR apps. Cog-1 in particular confused P1, respectively, with P1 elaborating: \textit{``When I think of like a flickering image, I think of a number of things. ... If I [had] the context of like, `Hey, this is for seizures', I'd be like, oh, yeah, like that. ... I interpreted [this guideline] as more aesthetic direction than, you know, you could cause somebody to fall out, hit their head and seriously injure themselves.''}
% \subsubsection{Understanding 3D Guideline Applicability in XR}
\label{understanding}
 %However, the applicability of these guidelines varied from a number of factors. %particularly due to considerations associated with XR hardware.
%Many of the most intuitive guidelines for developers were those that matched best design practices for XR. %\textbf{Colorblindness} was the most common impairment addressed by developers, with seven participants addressing colorblindness in their work. At least three %(P4, P20, P23)
% of the seven visual guideline evaluators had already incorporated Vis-1 in their work, and four participants suggested implementing Vis-1 by merely combining color and shape (P6, P9, P15, P23), a very XR-applicable guideline. All guidelines pertaining to \textbf{multimodality} (Vis-1, Cog-2, SH-1, SH-3) also seemed intuitive to participants. Similar to colorblindness, multimodality was seen as an a11y option benefiting those with and without a11y needs; particularly, P1 mentioned Cog-2 as \textit{``super essential''} and \textit{``one of the keys to VR development''} given that people may not immediately understand how to interact with the experience.
% P22 also indicated SH-1 as intuitive to people without a11y needs in noisy or crowded environments, given that they can focus on one task while staying informed about background tasks: \textit{``It's a good practice to have, because if they don't want to read or if listening is easier, or if at that point, they're looking at something else, they can still hear the main idea.''} 
However, some guidelines seemed less applicable to XR apps. \textbf{Timing} (Cog-4) for example seemed \textit{possible} to implement but not applicable to many XR apps, as real-time interactions (e.g., P20's firefighting simulator) seemed necessary to maintain immersion (\S\ref{balancinga11y}). Developers also argue this guideline would severely impact multiplayer apps, with additional design considerations needed to adjust fairness among players (e.g., Beat Saber~\cite{beatsaber} reducing players' score for playing at a slower speed). To best implement a range of appropriate speeds, P8 recommended developing \textit{``toward your extremes, low and high, and then everything else will kind of modulate in place.''}

\subsubsection{Guideline Implementation Timing and Responsibilities}
\label{a11ytiming}
Most participants agreed that a11y considerations should occur as early as possible, with \textbf{every} developer (25 out of 25) specifically mentioning that at least one guideline should be considered at the design phase, early prototyping stage, or up to a designer to implement.

% These slightly varied per guideline, but in most cases 

\textbf{\textit{When should a11y be implemented?}}
From our interviews, we identified six phases of development: early ideation, the design phase, early development, the development phase, testing, and post hoc.%~\daniel{could be confusing with our participant numbering scheme}. 
Participants overwhelmingly felt a11y features should be considered and implemented at the design phase (12 participants) or project start (4), with three participants indicating early development (P9, P11, P15), and only one at the development phase (P10 for SH-1, SH-2), testing (P5 for Mot-4), or post hoc (P5 for Mot-3). This timing could shift based on the size of the project; as someone also assuming design responsibilities, P5 emphasized a need for a11y implementation in the design phase for big tech projects, but a11y may be implemented post-hoc for freelance projects after feedback. Participants also noted certain guidelines as particularly important to develop as early as possible, with guidelines like Mot-5, Vis-2, and Vis-5 becoming ``prohibitively expensive'' late in development. As P6 put Vis-5, \textit{``Trying to patch it on after the fact sounds like a disaster.''}

\textbf{\textit{Who should implement a11y features?}}
Similarly, participants mostly agreed (18 participants) that \textit{designers} should be primarily responsible for implementing or enforcing accessibility features, with every guideline cited as a designer's responsibility or should be considered during the design phase by at least one participant (including by participants that self-identified as designers). %(though Mot-3 should still be considered during the design phase). 
In some cases, designers should be less involved in favor of developers (P19 for SH-2) or direct feedback from testers (P18 for SH-3).
%P1 P8 P4 P5 12 15 16 18 19
% Designer (Cog-1, Cog-2, Cog-3, Cog-4, Cog-5, Mot-1, Mot-2, Mot-4, Mot-5, Vis-1, Vis-2, Vis-3, Vis-4, Vis-5, P1, P5, P8, P13, P16, P21, P22, P23, P15)
% UX Designer (SH-1, SH-2, SH-3, SH-5, Cog-2, P12, P18, P19, P4, P20)
% Sound designer (SH-4, P12)
% UI designer (SH-4, Cog-2, Cog-3, Vis-1, Vis-2, Vis-5, P1, P2, P18, P19, P4, P9, P15, P24)
% Designer less involved (SH-2, P19) (SH-3, P18)
% SH-3:designers overlook what users think is important (P18)
% Technical designer (P22)
% Narrative designer (Cog-3, P1)
Fourteen participants agreed that \textit{developers} should be primarily responsible for implementing accessibility features, with Mot-4, all five SH guidelines, and four out of five Cog and Vis guidelines (not Cog-3 nor Vis-3) covered in these responsibilities. P3 in particular indicated that developers should do everything, no matter the guidelines asked, as standard for their big tech MR team. Additional responsibilities mentioned include artists (P1 and P4 for Cog-1 and Cog-5), the project manager (P16 for Cog-2), creative director (P2 for Vis-2 and Vis-4), or testers (P5 and P24 for Mot-4 and Vis-2).
Some developers like P8 mentioned responsibilities lie halfway on the developer and halfway on the designer, but since nine of our participants had both design and development roles, feature implementation may still be managed by the same individual. %P3 echoed that for their big tech mixed reality team, \textit{``It's just developers that do everything,''} noting their teams are solely composed of developers and project managers. 

% Another guideline that is hard to apply to XR is Vis-2 (UI resizing). Some developers noted that UI scaling would be implementable (P17, P23) with some caveats like pagination (P23), just needing to iron out design considerations. However, P2 and P17 both noted supporting UI \textit{reorganization} would be a \textit{``huge nightmare''} in XR (P2), potentially requiring a \textit{``recreation [of] the entire GUI system itself''} (P17). 
% Remaining 
% Developer (SH-1, SH-2, SH-3, SH-4, SH-5, Cog-1, Cog-2, Cog-4, Cog-5, Mot-4, Vis-1, Vis-4, Vis-5, P1, P2, P3, P4, P5, P12, P19, P8, P16, P20, P22, P23, P10) -- P3: devs do everything (SH-4 and SH-5 assumed for P3 given this statement)
% UI Developer (Vis-2, P24)
% UX Developer (Vis-2, P2)

% Artist (Cog-1, P1, P4)
% Tech artist (Cog-1, P4)
% VFX artist (Cog-1, Cog-5, P4)
% Animator (Cog-1, P1)

% PM (P16)
% Creative director (Vis-2, Vis-4, P2)

% Tester (Mot-4, Vis-2, P5, P24)

\textbf{\textit{Who should enforce compliance with a11y features?}}
Developers generally preferred XR-specific automation tools be implemented due to their use in 2D work (11 participants). % 1 3 4 12 15 16 17 19 20 21 22
P2 and P22 also recommended that the project manager (P2 for Vis-5), technical designer, or technical lead (P22 for Cog-1) manually review projects for a11y compliance should automated tools be insufficient. %Other developers like P3 noted the use of automated a11y tools in their 2D work and a need for XR-specific automation tools.

\textbf{\textit{When should a11y be shown to users?}}
Developers largely agreed that a11y should be shown to users at first time setup, not buried behind option menus, %supporting past research~\cite{citation needed}. 
As P1 outlines, \textit{``Having people dig through menus, like, you're only really inclined to do that when you're experiencing the discomfort already''} (P1), which isn't reasonable for Cog-1 if it directly affects a life-threatening condition (e.g., epileptic seizures). 
As a simple interaction default for subtitle customization in SH-5, P18 would offer participants a simple option on startup---\textit{``Can you see this text clearly? And if it is, click Yes.''}
P22 also suggested an a11y rating in the XR platform's app store, similar to Meta's comfort rating in the quest store.

% \subsubsection{Importance and Ease of Implementation of each guideline}
% \dk{Spreadsheet in comment here. Planning to make this section a bar chart/graph, but deprioritizing filling it out before Thursday or we cut this section entirely}
% https://docs.google.com/spreadsheets/d/1FzPKGo1aWD2S4hrBNbjXdFPDph1CRHN_Mq3rwQoWHGU/edit?gid=585667835#gid=585667835

\textbf{\textit{How should we prioritize guidelines?}} %\dk{this would be better presented as a bar graph imo as importance and ease of implementation. see overleaf comment here, above, that I won't get to before the deadline today. plan to include for R&R}
Developers rated guidelines as having varying importance, with those like Cog-1 rated as ``vitally important'' by P8: \textit{``It's literally everything. Because if you have a bad experience like this, number one, it could physically hurt somebody.''} Other guidelines like SH-5 seemed up to the platform to implement (P3), not the developers, with developers just passing through preferences from the user's settings.

\subsection{What XR A11y Support Methods do XR Developers Prefer?}
% Tool Preferences, i.e., Desired XR A11y Development Support Methods
\label{findings-support}
%\tochange{YZ: better connect to section 4.1 if possible, requested by R2 -- DK: may not be possible concisely? but let me know if you have any particular suggestions?} 
Prior a11y tools have focused on end-user experiences, but our findings highlight a critical need for developer-centered support tools that integrate with existing XR development workflows. Furthermore, developers used a variety of support methods at various stages of development. We provide the following framework to guide a11y tool development: A downloadable, free of charge, open-source 3rd party a11y package or set of reusable components is most preferred by developers (21 participants), however many considerations need to occur before organizations are willing to incorporate them into their projects. 
% p1 2 3 4 5 6 8 10 16 11 12 20 18 13 17 19 21 22 25 14 15

% \dk{immediately put a subsubsection on considerations here now ?}

%\subsubsection{Existing use of 3rd party support methods}
%We offer considerations towards developers' current use of 3rd party support methods that researchers should leverage when creating a11y support tools. %Besides base toolkits, most participants followed similar practices when developing their XR projects. 
%In this section, we outline some of the existing uses of toolkits to ensure compatibility with a 3rd party support method like a package or assets repository. %We use Figure~\ref{fig:supports} to visually demonstrate use of XR a11y in popular apps, alongside developer support methods for debugging, installing packages, and modifying parameters from in-app.
% Developers argue support tools should be easy to add to their projects (11 participants), ensuring that packages or components are easily . 
%p3 p9 p10 p13 p16 p17 p19 p20 p21 p22 p25

\subsubsection{Platform-Compatible A11y Toolkits}
%Most developers used platform-specific SDKs and toolkits based on the development platform they're using and the target device they're developing for. 
Understandably, we found that developers prefer a11y support that integrates with their existing development environments. Many participants used a platform-specific toolkit for XR development, with a nearly even split between OpenXR (7 participants), Unity XR Interaction Toolkit (6), and Microsoft's Mixed Reality Toolkit (MTRK) (6). Any a11y support tool should be compatible with these frameworks.

Despite differences in the toolkits used, developers highlighted a desire for compatibility between multiple common XR devices. P8 praised OpenXR, as it enables them to ensure device compatibility without the ``nightmare'' of designing the same feature for each platform from scratch. %\textit{``[OpenXR] really takes a lot of the backend handling of setting up left hand player, the right hand player, you know ... I just think without them, it's gonna run differently on every single project that makes any sense, right, which is something we don't want''} (P8). 
% Participants P19 and P22 also referenced OpenXR as a standard to target if an a11y support method did interface with a particular toolkit.

\subsubsection{Concerns Against Outside Direct Development Assistance}
% \tochange{Better intro sentence}
Many developers have concerns against outside influence on their code unless strongly compelled to, wanting to maintain control of their software, with some exceptions for freelancers and those with existing mod experience. %\tochange{DK: I'm a little worried this will unintentionally argue against our recommendation for a third party drag and drop package....how to rephrase? --- could soften the claim, not all participants; just say some participants have concerns...  }
% \subsubsection{Concerns against 3rd party packages}

\textbf{\textit{Third Party Packages.}}
Although many developers leveraged 3rd party components and libraries (11 participants), developer support methods like a11y packages for XR platforms often require specific incentives for developers to adopt based on company size. These packages are often quickly disregarded by some startups (P2, P12) and need to be cleared by big tech legal teams, even if open source.
%Many participants (e.g. P2, P3, P8) demonstrated their practices founded from a web development background, with access to an abundance of 3rd party libraries. \textit{``At least for JavaScript packages people use drag and drop packages, open source packages all the time. And [company] as well, as we're not, we're not anti-open source; if we find something that's useful, we'll use it''} (P3). P3 also mentioned potential licensing issues unless the package grants appropriate permission for use (e.g. MIT), otherwise needing to ask their company's legal team. %Citing startups' ``survival mode'' mentality, P23 mentioned a startup developer's thoughts when quickly evaluating a support method:
%\textit{``Does this solve my problems? No? Throw it in the trash.''} 
%Developers like P2 and P12 were skeptical of their use. 
P2 mentioned startup developers having \textit{``not invented here syndrome''}---developers at these companies feel personal attachment to their projects, with a philosophy that they \textit{``[already] know everything,''} not wanting to involve consultants nor code written outside their organization. Other startups may be more likely to pick up 3rd party tools, quickly evaluating their ``worth'' (P23) towards speeding up their development time (P19). Big tech companies also seem to be more hesitant to use 3rd party packages, preferring to use internal tools if they exist, or requiring packages that are open source if they do not have an internal equivalent. Some big tech developers even offered to rewrite custom versions of packages to only include the features they need, citing build time as a need for lightweight packages (P15). 

\textbf{\textit{Game Modification.}}
Developers, particularly those at midsize and big tech companies, generally disapprove of adding 3rd party game modification (``mod'') support, i.e., allowing community members to modify their games, %interface methods allowing people to independently modify their games. %One may believe that since developers would want the easiest integration method possible with the highest payoff, they may be more open to 3rd party game mods as this method shifts the a11y feature burden off developers and onto their community. 
due to concerns about security (P24), legal problems (P25), platform integration (P17), or unfair advantages in multiplayer (P1). Only five developers directly supported implementing mod integration into their apps, all of whom either have preexisting modding experience (P5) or outlined strict stipulations of the data that mods would be allowed to access---like exposed metadata parameters but not feature code (P22).

% \subsubsection{}
% \textbf{\textit{Automated A11y Checking Tools vs. Consulting.}}
% Most developers preferred an automated tool for a11y compliance checking (12 participants).
% Six developers appreciated personalized consulting support, either by a dedicated role at their company or by hiring an external service %(e.g., P20). % perhaps similar to sites like YouDescribe~\cite{youdescribe} that provide online platforms to speak with a11y consultants and for PWD to use. 
% Developers at P2's startup are very anti-consultant, coming from a  setting high standards for the support they could receive. 

% ----------------------------------------

%\subsubsection{Lacking Specific Guidance on Implementation Recommendations}
%Even with 3rd party support methods like an a11y package or toolkit, developers would also like targeted assistance in their projects through explicit examples of a11y solutions and optimal forms of implementation. 
% \subsubsection{Guidance} support includes more advisory support, including the current lack of standard XR guidelines and accessibility teams.
% \label{subsection:guidance}
% % advisory content, blog posts, general principles

\subsubsection{Feature \& Code Examples.}
Developers want specific examples of a11y features to implement. These examples should include the ``best'' implementation for each a11y feature (P15) and be presented in a manner that's ``easily digestible'': e.g., providing both accessible and inaccessible examples (e.g., safe and unsafe flash rates in Cog-1, with a warning presented over the unsafe option); videos or GIFs of PWD unable to use an app until the support method is added; % and able to use the app once the feature is added %contrasting a video or GIF with a PWD able to use the same app with the a11y feature
and code snippets, including downloadable projects and Unity scenes. 
%\textit{\textbf{Blog posts.}}
Four developers %P2, P9, P11, and P24 
preferred to view a11y feature examples or support in the form of blog posts, with each post containing best implementations, descriptions of performance costs, and relative importance at each phase of development. Six developers would also appreciate personalized consulting support, either by a dedicated role at their company or by hiring an external service. % check PX on where these should be published, e.g. P2's HackerNews type thing or another participants' suggestion like Unity forums / big tech publicizing / who ? 
% P2: \textit{``If you had that spelled out in a blog post, there'd be more developers that me and then maybe you would understand that more optimized way to do it, then I would do it first path.''}

% ----------------------------------------

% \subsubsection{Execution} support includes technical support methods recommended by developers, including directly implementable accessibility solutions into their projects.
% \label{subsection:execution}
% hands-on implementations, technical solutions

\subsubsection{Developers Prefer Directly Implementable Solutions}
Ultimately, we found that developers want an open source, drag \& droppable package that enables them to directly implement XR guideline-compliant a11y features. This package should be well-documented (9 participants), easy to use (P3, P16), have its own UI (P19), and be able to automatically verify compliance with accessibility standards (P22). Most developers did prefer an automated tool for a11y compliance checking (12 participants). %, providing warnings or errors (P22). %in their platform's developer console (P22). 
This package may provide a skeleton structure for XR apps (P5) but should be functional at all stages of development.
% P23 provided an overview of the needs for an approachable a11y package or toolkit for XR, uniquely noting concerns associated with the time it takes to implement the toolkit itself, even if the toolkit incorporates many accessibility features:
P23 uniquely emphasized that successful accessibility tools must minimize implementation time, even if the toolkit incorporates many accessibility features:

% \tochange{DK: I do still think this is a great quote but don't know how to shorten since it just takes up a lot of space now (and even more when double-column) ---it's fine}
\begin{quote}
\textit{``\textbf{You think the tool is providing you the time to implement it.} Good software tools are very minimal in terms of their implementation time. If a tool takes 25 hours to implement into your application, that's misery. If it takes 25 minutes, that's an actually good tool---a tool needs to be able to be picked up and used quickly. ... %So XR toolkit[s] %stuff
%would need to be very immediately approachable, and very usable. So 
% ``... Your guideline descriptions and stuff needs to be like, popcorn easy to consume, easy to understand. ... 
You're constantly fighting an attention warfare; having good visual content that draws them in and immediately communicates the ideas, ... putting some images in there is going to be really helpful for getting people to overcome their initial attention of like, oh, I should pay more attention to this. Okay, I will watch you know, the tutorial video on how to implement the toolkit or oh, here's the link to the toolkit SDK docs.''} 
\end{quote}

\section{Discussion}
Our study reveals the unique challenges and opportunities in XR that fundamentally reshape accessibility decisions across different organizational contexts. In this work, we interviewed 25 XR developers from four types of organizations, ranging from freelancers and startups to employees at XR platforms and big tech companies, developing a comprehensive understanding of their practices, attitudes, and challenges towards developing accessibility features in their apps. We answer our three research questions as follows:
 % \item[RQ1.] What challenges do XR developers face against developing XR accessibility features?
Firstly, developers face challenges incorporating accessibility features without sacrificing visual effects and interactions that they believe make their app immersive (\S~\ref{balancinga11y}). As XR applications utilize unique interaction methods, developers have difficulty onboarding first-time users (\S~\ref{onboarding}), resolving input (\S~\ref{io}) and performance considerations (\S~\ref{performance}) associated with some a11y features, and evaluating the time/cost necessary to implement a11y features (\S~\ref{motivations}; RQ1). Secondly, a lack of formal XR training (\S~\ref{lack-training}) combined with a multitude of responsibilities (freelance, startups) or lack of control over what you implement (midsize, big tech) leads developers away from implementing a11y features (\S~\ref{motivations}). Practices of reusing code would make future a11y integration easier after the first project (\S~\ref{reusing}), but starting a11y feature design is challenging, particularly if the initial prototyping goal conflicts with accessible design (RQ2).
 % \item[RQ2.] What current XR development practices do developers follow? How conducive are they to implementing XR accessibility features?
Lastly, we evaluated state-of-the-art guidelines for 3D virtual worlds and applied them to XR, determining that guidelines need more XR-specific context to be understood (\S~\ref{understanding}) and developers need sufficient motivation to implement them (\S~\ref{motivations}). Developers also desire easily implementable, open source toolkits to help them implement a11y features (\S~\ref{findings-support}; RQ3).
 % \item[RQ3.] How applicable are guidelines? What support methods would best enable XR developers to implement XR accessibility features?
In this section we discuss implications for creating effective XR a11y support methods informed by industry XR developers. We also discuss limitations of this work and opportunities for future work, including developing the support methods offered.
 
% ----------------------------------------
%
%          Subsection break :)
%
% ----------------------------------------

\subsection{Systemic Deprioritization of Accessibility Integration}
%A11y Support Not Prioritized Due to Ableism
% As mentioned in \S~\ref{findings-practices}, most developers followed similar development practices in their projects, but we did see some differences in developers' use of 3rd party plugins. 
%\tochange{YZ: based on 2AC's review, here we need to summarize and highlight how developers conceptualize disabilities and power dynamics in workplace (e.g., priority of the company's business goal limit a11y)}
Our findings reveal how existing development practices overlook a11y as a core requirement of XR development. Developers we interviewed noted a dependency on big tech companies to publicize a11y features, with many celebrated a11y examples being AAA titles like God of War: Ragnar{\"o}k~\cite{godofwar} and The Last of Us Part II~\cite{lastofus}. Many startups in ``survival mode'' believe they lack the ability to compete with these projects, and those that do rely on post-launch feedback rather than integrating a11y features from the start.
% vampire survivor and loop hero are games with limit effects, one handed mode,  
%\tochange{DK: R1 mentioned some additional indie games but they're 2D action games rather than 3D VR experiences. Deciding not to include. We do talk about some challenges in the next section though and ref 4.2 if that's helpful (even if not 4.1 as R1 described). --- can you find a place to insert them in RW? no description, just add a ref somewhere}
Interestingly, some XR-specific game studios like Owlchemy Labs have successfully implemented a11y features in games like Job Simulator and Cosmonious High (P1, P24)~\cite{owlchemylabsa11y}%and sharing expertise with their community}%while they maintain accessibility documentation and }%\change{Many indie games particularly in 2D like Vampire Survivors and  feature accessibility options like one-handed mode and disabling flashing visuals, but %and websites like \textit{Can I Play That?} publish accessibility reviews in games.
%We highlight a need to further amplify XR titles that implement accessible features, not just th
%studios like Whitethorn Games and Owlchemy Labs in particular have been championed by some gaming news outlets (e.g., \cite{whitethorna11yblind, indiestudiospioneeringa11y}) for their accessibility , 
---%However, AAA titles are likely hailed as such because of reasons our participants indicated---they're notable titles that people recognize, encouraging smaller studios to follow suit. 
but these studios are not the norm. Our participants point to the most popular VR applications like Beat Saber, Rec Room, and VRChat (P2, P18) as the trendsetters for a11y implementation. %\dk{prev. sentence sounds a bit off but addresses R1's \textbf{\textit{new}} Owlchemy complaint} %(Figure ~\ref{x}) 
% Solutions can also emerge from experiencing VR sources; P16 for example emulated tunnel vision motion sickness support from Skyrim VR~\cite{wilhelm2017reimagining} into their toolkit. %demonstrating %\change{that users simply trying an accessible feature may inspire them to emulate it}. %---also making the app as a whole more comfortable for all users. 
As a whole, developers systemically deprioritizing a11y features reveals a structural challenge in XR development, where business pressure limits a11y development and creates a false divide, discounting inclusivity for profit---while in reality a11y feature development improves the experience for all users \cite{microsoftInclusive101}. %\S~\ref{deva11yresponsibility}). Transforming this system requires reconceptualizing accessibility not as an add-on, but as a fundamental aspect of XR development, acknowledging our responsibility to create truly inclusive digital experiences.

\subsection{Developing Accessible XR Solutions and their Benefactors}

\subsubsection{Implementing A11y Throughout the Development Lifecycle}
% \subsubsection{Developers' Diverse Responsibilities.}
Our findings demonstrate how XR's unique development practices complicate a11y implementation across different organizational contexts. Despite evaluating many company sizes, we found most of our developers cited similar experiences in their XR development practices, having self-taught XR development and holding many responsibilities within their swiftly-moving teams. Alongside developers, many of our participants were executives of their companies or asset and interaction designers, fulfilling roles that encompass the entire software development lifecycle. %\anonymous{Since many of our 'developers' were also designers and managers we briefly address our non-engineering stakeholders here.} 
We mention in \S\ref{a11ytiming} the need for a11y implementation as early in the development process as possible. However, participants' diverse responsibilities outside of pure development (\S\ref{responsibilities}) leaves little time to develop a11y features nor test with PWD (\S\ref{testingpwd}). Freelancers and startup developers believe their workplaces would be less receptive to a11y development until after their app releases (e.g., P12, P16), even if it's more difficult (P4), unless their client specifically requests it (e.g., P1, P5, P20--P22). %Therefore, pushing against workplace power dynamics may be more difficult.
Midsize and big tech developers \textit{including designers} (e.g., P15) appear more receptive to urgently improving their XR apps' a11y at the start of development, but due to the scale of big tech projects, participants may not want to add more features if the project has already scaled past a terminal point (e.g., P9). Since developers believe guidelines like Vis-5 become prohibitively expensive late in development, they encourage startup developers to start as soon as possible (P6); however, guidelines should also be formatted in a way that implementing them does not become an insurmountable task. 

Based on these findings, we maintain our recommendation for drag-and-drop toolkits, which may significantly \textit{reduce} overhead while encouraging early a11y feature implementation---allowing even late-in-development projects to effectively include a11y features without overwhelming developers. We also note that although platforms like Babylon Native enable unique benefits like device-native screenreader integration~\ref{nativedevicehooks}, we recommend prioritizing efforts for future drag-and-droppable a11y toolkits towards Unity, given its widespread adoption (used by 23 of our 25 participants) and its position as a ``go-to tool for creating XR applications in practice''~\cite{nebeling2022tools}.

% \tochange{add workplace power dynamics mention and refer back to \S2.1} \tochange{YZ: also talk about that there is no clear responsibility assignment for a11y in the emerging XR industry, so that no one wants to take that responsibility. R2: "The technical content would benefit from more explicit connection to development processes. While the paper notes that "developers kicked accessibility responsibilities to designers" (p.17), it doesn't fully explore the implications for workflow integration."
 
\subsubsection{Need for Automated Verification Tools}
One frequently requested feature for a11y support was an automatic scene scanner for a11y compliance. However, the implementation of such a feature may prove challenging for developers that dynamically generate their scenes---we found that at least five developers dynamically created objects in their scene (P8, P14, P15, P18, P22). %(P8, P14, P15, P18, P22). 
This practice would likely make some a11y checking impossible before runtime, such as z-fighting (P4) or performance limitations (P21) potentially violating Cog-1. %,.
Support methods like an XR a11y toolkit are very important for people lacking a11y resources, enabling significantly more a11y consideration in XR apps (P25). Toolkits that fit developers' recommendations outside of XR a11y do exist, like Sarmah et al.'s \textit{Geno}~\cite{sarmah2020geno}, which provided a high-level interface to implement voice commands into web applications.
Developers indicated that support systems should be able to take one step further and recognize programs like screen readers on one's XR system and import their settings.
% take that a step further and recognize standard a11y system, like NVDA equivalents running on your device. 

% Could use something like Sarmah et al.'s Geno \cite{sarmah2020geno}, a tool for integrating voice inputs into web applications. As implementation of voice commands would typically require significant effort and knowledge from the developer, Geno automated a majority of the process to instead provide a high level interface that even developers with little expertise could utilize.

% \subsubsection{Need for Accessible Design Solutions for All}
% Echoing past research~\cite{dudley2023inclusive, persson2015universal}, we find that accessible design solutions benefit both people with impairments and general users. With consumer AR- and AI-powered smartglasses like Brilliant Labs Frames~\cite{bLabs}, XREAL Air~\cite{xreal}, and RayBan | Meta smartglasses~\cite{RayBanMeta} entering the market, accessible solutions need to be developed from the start to maximize usability in- and outside of people with disabilities. 

\subsubsection{Maintaining Tools Over Time.}
%DK: This was also a question brought up by the IMWUT reviewers; do you have any other ideas besides open-sourcing on github, which follows other recommendations? Are there any other papers can I cite that we can point to that say the same thing? (SeeingVR does not) --- not anything on top of my mind; but 

Another concern mentioned by developers (P7, P9) that we share is a lack of maintenance of support methods over time. Prior work in computer science theory and software engineering emphasizes a need for modular, easy to read code; clear and quality documentation; dedicated code reviewers; and an active todo list to motivate continued contribution to the project~\cite{aberdour2007achieving, crowston2008free, greenberg2007toolkits, guimaraes2013life}. Additional programs like the XR Association~\cite{xrassociation} and XRAccess~\cite{xraccess} that aggregate XR a11y development resources (e.g., ~\cite{xrAccess2024github}) may help publicize toolkits for continued development, but this would require a11y research systems to open-source their code for further iteration or re-implementation (unlike ~\cite{harada2007voicedraw, ji2022vrbubble, nair2021navstick}). Accessible systems %particularly developed in commonplace applications like Unity 
should consider establishing sustainable maintenance frameworks that balance code quality and 3rd party collaboration opportunities. Projects like SeeingVR~\cite{seeingVRtoolkitgithub} require submitting an additional license agreement to contribute, which may deter developers from submitting their changes. Given that at least 18 of our participants were already familiar with established software engineering version control practices or platforms like GitHub, we believe such systems could be made available for reuse by the research community and include clear contribution guidelines that support external maintenance. Ultimately, reducing these contribution barriers while encouraging standard practices may be vital to extending accessibility tools' lifespan.

% reducing barriers to contribute while maintaining established software engineering code review standards may be vital to extend accessibility tools' lifespan. Such systems should be available for reuse by the research community and include clear contribution guidelines that support external maintenance. % on the toolkit
%In their work on toolkit development Greenberg recommends such reusable components over time helps toolkits evolve once disseminated unto the community~\cite{}}.

% to change: 
% - some community/government support to enable creators of 3rd party a11y tools to continue working on the tool could be helpful; most tools or software won't allow others to contribute to their codebase}
% - .. contributing by 3rd party is very tricky and definitely needs to be reviewed and screened
% - need to soften this claim --- "these contribution steps should be as simple as opening a pull request with changes"

\subsubsection{Exposing Data for A11y}
Prior a11y tools have attempted to modify applications after development, but our findings suggest the need for standardized data exposure mechanisms between applications and a11y services. Developers expressed openness to exposing a11y metadata but concerns about security and performance; %However, we recognize that end-user XR application developers are rather protective of their projects, so 
providing direct access to applications' files in a manner similar to 3rd party game modifications is unlikely. 
Past systems like SeeingVR~\cite{zhao2019seeingvr} and RealityCheck~\cite{hartmann2019realitycheck}
implement post hoc a11y plugins by ``hacking in'' to projects, but developers expressed preferences for packages or toolkits %Developers should expect conventions similar to Unity defaults; however, we cannot predict this behavior for projects with procedurally generated content as we do not have access to metadata not exposed by the project---particularly if procedurally generated scene graphs become industry standard due to their perceived scalability. Furthermore, if the end goal of a research project is for mass adoption of the system in industry, we find it unlikely that developers would install post hoc methods at the sacrifice of their perceived security, even if it would benefit a11y as a whole.
% We do also note concerns that developers late in the development cycle would become discouraged by increased performance costs, choosing to continue avoiding a11y implementation as a result.
%developers appeared to prefer a package or toolkit 
similarly structured to SeeingVR, i.e., in the form of a platform-specific package that adds a11y features through a central manager~\cite{zhao2019seeingvr}. %(in SeeingVR's case, a GameObject template prefab which automatically added all tools to the VR app and controlled through their Accessibility Manager~\cite{zhao2019seeingvr}). 
Additionally, developers \textit{did} seem open to exposing particular metadata from their projects to external programs, much like 2D screenreaders like NVDA and JAWS read from alt text labels on websites for images, but this practice may require more work from developers and a separation between the emitting program (e.g., a Unity package) and a receiving program (e.g., a system-level XR screenreader).

\subsubsection{Engaging PWD in XR A11y Teams}
%change the title to something like: engaging PWD in XR A11y solutions. --- since all reviewers asked about explicitly discussing how to better involve in the current development flow" and 

% As Unity was by far the most used platform ($n=23$, excluding P6 and P17), we saw a variety of other platforms used, including several webXR platforms and two custom internal engines. Other XR development platforms such as Unreal Engine ($n=11$) and Godot (P22) were mentioned but not as widely used. Participants making webXR apps largely used web-based development tools like Glitch (P6, P9), BabylonJS Playground (P3), and PlayCanvas (P14). Two developers used custom engines proprietary to their organization (P8) or target hardware (P25).

% \tochange{refine this section to focus more on engaging PWD explicitly. Currently, things seem to be mixed, suggesting providing more knowledge to developers that can be done in both guidelines, direct feedback from PWD, and others...but we need to highlight the engagement of PWD more. either do it here, or add a different subsection.}
While prior work has demonstrated the value of inclusive design, our findings reveal systematic exclusion of people with disabilities from XR development processes, limiting the effectiveness of accessibility solutions. Addressing this gap requires fundamental changes to development team composition and power dynamics.
A significant barrier to a11y implementation is developers' lack of knowledge about a11y requirements, with developers appearing more willing to incorporate accessible features if they can quantify their input costs (time/monetary cost) against measurable output (increased userbase, positive reviews, and lack of legal disputes) (\S~\ref{motivations}). This transactional approach does not recognize the ethical importance of a11y and perpetuates a cycle of exclusion.

Compounding this issue, systemically excluding PWD from XR development perpetuates existing stigma and barriers to a11y. While developers like P14 cited a ``lack of access to testers with disabilities'', this comment implies a deeper issue: PWD are not consulted in the development process, despite designers often overlooking what their users think is important (P18).
Engagement with PWD must establish PWD as equal stakeholders in development teams; however, technical integration alone is insufficient---we must recognize how disability intersects with other aspects of identity such as race, gender, class, and sexuality to inform accessible design~\cite{costanza2018design,gauthereau2020tagged,dudley2023inclusive}. %and address solutions that consider their diverse needs~\cite{}. This approach not only improves accessibility but challenges the traditional power structures that have historically excluded PWD from technological development. \dk{earmarking as something from reviews to potentially change but also sounds fine to me}

\subsection{Limitations and Future Work} %and Improvements in XR A11y Guidelines Since Interviews}

Our study has certain limitations. Although we tried our best to encompass a diverse range of XR organizations, the current XR industry is so fragmented that almost every company has their own practices covering a large variety of types of apps. A survey targeting a broader range of developers may yield different perspectives on accessibility in XR, providing new insights into guidelines and tool design. Additionally, as we attempted to equalize the coverage of guideline discussion between participants, we may have chosen guidelines more or less familiar to developers or relevant to the XR apps they create. 
We acknowledge the possibility of recruitment bias, since our topic potentially attracted a higher proportion of developers already supportive of accessibility. Although our study focused on developers, future work may expand to non-engineering roles and stakeholders, including project owners and clients. We also acknowledge that non-3D a11y guidelines could be expanded and applied to XR; we %Our guideline evaluation focuses on XR-relevant guidelines. 
recommend future work explore this connection. %\tochange{YZ: also mention the grouping issue, speech and hearing.}. 

\section{Conclusion}
We present our insights from interviewing 25 XR developers across a range of organizations about their perspectives and practices towards accessibility. We presented them with existing accessibility guidelines and discussed their implementation of them, as well as the tools they would need to implement them into their projects. Our findings provide potential strategies for enhancing the feasibility of accessibility implementation across the XR industry. % , notably, ?
Many participants demonstrated their practices founded from a software development background, demonstrating a need to develop new solutions that are specifically targeted towards XR applications. We found that developer support methods for XR developers include written materials like guidelines, but most developers would prefer platform-targeted, open-source packages they can drag-and-drop into their projects or pull specific features from. Further research is needed to establish robust accessibility standards that provide more comprehensive developer support. Additionally, the ease of incorporating people with disabilities, particularly from the design phase, needs to be better conveyed to XR developers, and to demonstrate a ``worth'' for developers to benefit a large number of users. After all, if \textit{``XR is already an answer for an accessibility issue''} (P21), why create new barriers?

% \begin{acks}

% \end{acks}

%TC:ignore
\printbibliography
%TC:endignore

\appendix
\section{Themes}
\label{appendix:themes}
%TC:ignore
\begin{table*}[ht!]
  \small
  \caption{Non-Exhaustive Summary of Themes, Sub-themes, and Corresponding Example Codes} %\anonymous{A full list of codes formatted as a LaTeX table would be impossible to include, but we've included a subset of codes in this table.}
  \Description{A table summarizing non-exhaustive themes and sub-themes in XR accessibility development. From left to right, column titles are Themes, Sub-Themes, and Example Codes. The table contains 6 main themes: Difficulties of Feature Integration, Current Development Practices, Motivations, Developers' Current XR A11y Solutions, Applicability of 3D Virtual World Guidelines to XR, and Preferred XR A11y Support Methods. Under "Difficulties of Feature Integration," sub-themes include Sacrificing A11y for Immersion, A11y Issues and Needs in XR Interactions, Disability Representation in XR, Performance Limitations of Standalone HMD's, and Onboarding First-Time Users. Example codes range from "balance a11y and immersion" to "tag objects as interactable." "Current Development Practices" includes sub-themes like Lack of Formal XR Training, Developers' Diverse Responsibilities, Dynamic Code Generation, Reusing Code Between Projects, and Testing with PWD. Example codes include "no formal XR training" and "PWD appear in general testing." The "Motivations" theme covers Motivations, Hindrances, and the idea that everyone has a responsibility to be an a11y developer. Example codes include "time," "money," "survival mode," and "accessible design." "Developers' Current XR A11y Solutions" encompasses Incorporating Alternate Input, Hooking into System A11y Support, and Examples of Existing A11y Solutions. Example codes mention "rewired," "custom hooks," and "colorblindness filters." "Applicability of 3D Virtual World Guidelines to XR" includes Overall Knowledge of Guidelines, Understanding Guidelines, and When to Implement a11y. Example codes reference "WCAG" and "a11y in design phase." The final theme, "Preferred XR A11y Support Methods," covers various sub-themes such as SDK's/Toolkits, 3rd Party Packages, Game Modification, Automated A11y Checking, Consulting, Provide Examples, and Need to be Directly Implementable. Example codes range from "toolkits" to "easy to implement."}
  \label{table:themes}
  \begin{tabular}{p{2.5cm}p{3.5cm}p{5.5cm}}
  \toprule
Themes & Sub-Themes & Example Codes \\
\midrule
\textbf{Difficulties of Feature Integration} & Sacrificing A11y for Immersion & balance a11y and immersion \\
\cmidrule{2-3}
 & A11y Issues and Needs in XR Interactions & text illegible due to resolution; need XR equivalent of ARIA label; AR small display; spatial audio for BLV; spatial audio difficult; AD from spatial audio location; captions; caption implementation difficult; prefer subtitles; locomotion challenging for PVI; thumbstick turning harms mental map \\
\cmidrule{2-3}
 & Disability Representation in XR & normative view \\
\cmidrule{2-3}
 & Performance Limitations of Standalone HMD's & vr performance considerations; a11y feature requires performance overhead; game store performance requirements \\
\cmidrule{2-3}
 & Onboarding First-Time Users & tag objects as interactable; video tutorial insufficient \\
\midrule
\textbf{Current Development Practices} & Lack of Formal XR Training & no formal XR training; on-the-job XR training; self-taught; online courses; online forums; personal projects; YouTube; Udemy; Udacity; documentation; Reddit; Discord \\
\cmidrule{2-3}
 & Developers' Diverse Reponsibilities & developers wear many hats \\
\cmidrule{2-3}
 & Dynamic Code Generation & code easier to merge vs prefabs \\
\cmidrule{2-3}
 & Reusing Code Between Projects & reusing code; reuse a11y subsequent projects \\
\cmidrule{2-3}
 & Testing with PWD & PWD appear in general testing; no PWD testing; direct user feedback from PWD \\
\midrule
\textbf{Motivations} & Motivations & time; money; number of users; reviews; legal \\
\cmidrule{2-3}
 & Hindrances & survival mode; MVP \\
\cmidrule{2-3}
 & Everyone has a responsibility to be an a11y developer & accessible design; also just best design practices; including accessible design lowers costs \\
\midrule
\textbf{Developers' Current XR A11y Solutions} & Incorporating Alternate Input & rewired; merge cube; xbox a11y controller \\
\cmidrule{2-3}
 & Hooking into System A11y Support & custom hooks; Unity lacks a11y hooks; hook into platform functionality \\
\cmidrule{2-3}
 & Examples of Existing A11y Solutions & colorblindness filters; one-handed mode; text size; UI magnification \\
\midrule
\textbf{Applicability of 3D Virtual World Guidelines to XR} & Overall Knowledge of Guidelines & current guidelines broad; WCAG; Microsoft's MR Accessibility Standards \\
\cmidrule{2-3}
 & Understanding Guidelines & not applicable for real-time gaming \\
\cmidrule{2-3}
 & When to Implement a11y & a11y in design phase \\
\midrule
\textbf{Preferred XR A11y Support Methods} & SDK's/Toolkits & toolkits; hardware-bound SDK; use platform-specific SDK's \\
\cmidrule{2-3}
 & 3rd Party Packages & use of 3rd party packages \\
\cmidrule{2-3}
 & Game Modification & modding security concerns; offload responsibility to modding community; multiplayer modding concerns \\
\cmidrule{2-3}
 & Automated A11y Checking & lacking automated a11y checks for 3D \\
\cmidrule{2-3}
 & Consulting & consulting \\
\cmidrule{2-3}
 & Provide Examples & provide examples; sample scenes; GIFs; videos \\
\cmidrule{2-3}
 & Need to be Directly Implementable & easy to implement; drag and droppable plugin \\
\bottomrule
\end{tabular}
\end{table*}
%TC:endignore
\clearpage
% \section{Prompt}
% \input{tables/apndx-prompt}
\section{Beat Saber Color Options}
%TC:ignore
\label{appendix:beatsaber}
\begin{figure}[h!]
    \centering
    \includegraphics[width=1\textwidth]{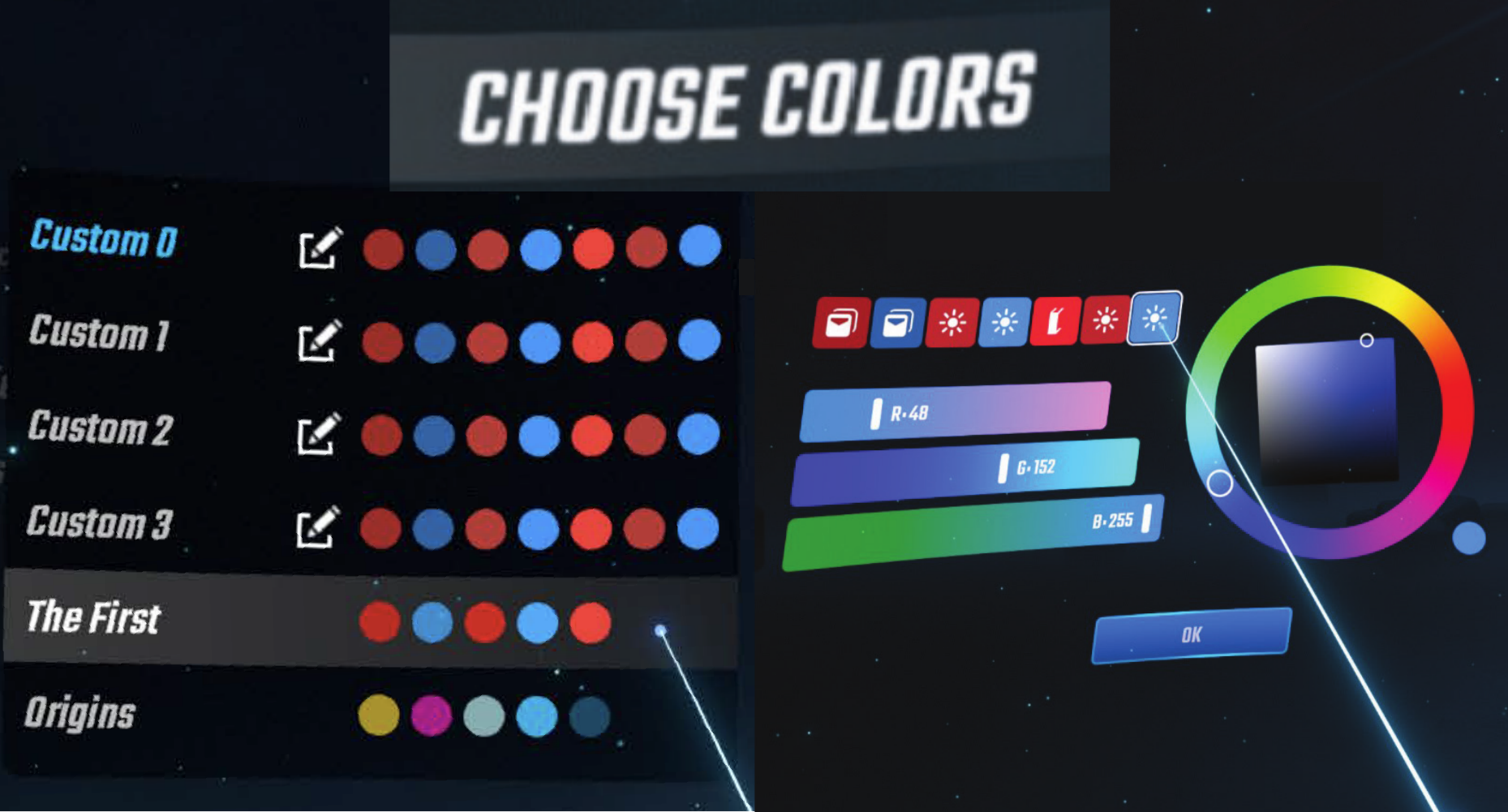}
    \makebox[\textwidth]{%
        \parbox{1\textwidth}{%
        \caption{We show an example from one of the most popular~\cite{faric2019players} VR games, Beat Saber~\cite{beatsaber}, which offers custom color settings for a variety of virtual objects. Selecting an object (Right) like notes, lights, or wall, then picking a color from the picker, will cause the respective virtual object to change to that color. Users can create up to four custom color palettes (Left, labeled Custom 0 through Custom 3) or choose from a variety of presets themed from in-game collections (e.g., ``The First'', ``Origins''). As of November 2024 there do not exist built-in presets for common color vision deficiencies (e.g., protanopia, deuteranopia, tritanopia)~\cite{colorblindnesstypes}.}
        }
    }
    \Description{Edited screenshots from the VR game Beat Saber displaying the in-game color picker. The top of the image reads "Choose Colors". The left half of the image shows six color palette options. The first four read "Custom zero" through "Custom three", and the last two read "The First" and "Origins". All six options are selectable. The four "custom" options are editable. Each color palette is shown with five to seven circles next to it, where every one except "Origins" shows a Red and Blue color scheme. Origins shows an orange, pink, and blue color scheme. On the right half of the image is a series of buttons next to a color picker, with a blue OK button at the bottom. Seven buttons on top show icons with two notes, two lights, one wall, then two more notes. each set of two alternates red then blue. Selecting one icon changes the color picker to that color. The color picker is a red, green, and blue slider on the left side, and a circular wheel on the right side. A square sits in the middle of the wheel to pick saturation that is colored in a gradient, with dark blue on the right side and gray on the left, white in the top left corner and black on the bottom. A virtual reality pointer points to the color palette option called "The First" on the left half of the figure, and another virtual reality pointer points to the rightmost blue light icon.}
\end{figure}

\end{document}